\documentclass[aps,pra,showpacs,superscriptaddress, table, twocolumn]{revtex4-1}

\usepackage[dvipsnames]{xcolor}
\usepackage{graphicx}
\usepackage{bm}       
\usepackage{amssymb}  
\usepackage{amsmath}  
\usepackage{subfigure} 
\usepackage{afterpage}
\usepackage{array}
\usepackage{multirow}
\usepackage{makecell}
\usepackage{hyperref} 
\usepackage{ifthen}
\usepackage{acro}

\usepackage[utf8]{inputenc}
\usepackage[english]{babel}

\DeclareAcronym{DFT}{
short = DFT ,
long = Density-Functional Theory
}

\DeclareAcronym{SOAP}{
short = SOAP ,
long = Smooth Overlap of Atomic Positions
}

\DeclareAcronym{ACE}{
short = ACE,
long = Atomic Cluster Expansion
}

\DeclareAcronym{MILAD}{
short = MILAD,
long = Moment Invariant Local Atomic Descriptors
}

\DeclareAcronym{RR}{
short = RR,
long =  Ridge Regression
}
\DeclareAcronym{KRR}{
short = KRR,
long = Kernel Ridge Regression
}

\DeclareAcronym{ACSF}{
short = ACSF,
long = Atom Centered Symmetry Functions
}

\DeclareAcronym{PCA}{
short = PCA,
long = Principal Component Analysis
}

\DeclareAcronym{GAP}{
short = GAP,
long = Gaussian Approximation Potential
}

\usepackage[normalem]{ulem}
\hypersetup{
    colorlinks=true,
    linkcolor=blue,
    citecolor=blue,      
    urlcolor=blue,
}

\usepackage{verbatim}

\newboolean{show_figs}

\begin{document}

\setboolean{show_figs}{True}

\title{Compressing local atomic neighbourhood descriptors}
\author{James P. Darby} \email{jpd47@cam.ac.uk}
\affiliation{Engineering Dept, University of Cambridge, Trumpington St, Cambridge CB2 1PZ, United Kingdom}
\author{James R. Kermode} 
\affiliation{Warwick Centre for Predictive Modelling, School of Engineering, University of Warwick, Coventry CV4 7AL, United Kingdom}
\author{G\'abor Cs\'anyi}
\affiliation{Engineering Dept, University of Cambridge, Trumpington St, Cambridge CB2 1PZ, United Kingdom}

\vskip 0.25cm

\date{\today}

\begin{abstract}
Many atomic descriptors are currently limited by their unfavourable scaling with the number of chemical elements $S$ e.g. the length of body-ordered descriptors, such as the \ac{SOAP} power spectrum (3-body) and the \ac{ACE} (multiple body-orders), scales as $(NS)^\nu$ where $\nu+1$ is the body-order and $N$ is the number of radial basis functions used in the density expansion. We introduce two distinct approaches which can be used to overcome this scaling for the \ac{SOAP} power spectrum. Firstly, we show that the power spectrum is amenable to lossless compression with respect to both $S$ and $N$, so that the descriptor length can be reduced from $\mathcal{O}(N^2S^2)$ to $\mathcal{O}\left(NS\right)$. Secondly, we introduce a generalized \ac{SOAP} kernel, where compression is achieved through the use of the total, element agnostic density, in combination with radial projection. The ideas used in the generalized kernel are equally applicably to any other body-ordered descriptors and we demonstrate this for the \ac{ACSF}. Finally, both compression approaches are shown to offer comparable performance to the original descriptor across a variety of numerical tests. 
\end{abstract}

\maketitle

\section{Introduction}
Increases in computational power have made it possible to use quantum mechanical methods \cite{burke2012perspective, nightingale1998quantum, bartlett2007coupled} to model materials at a level of accuracy where chemical transformations can be usefully studied. This ability has opened the door for \textit{in silico} materials modelling to usefully compliment, and sometimes even replace, the need for costly experiments. A major drawback of such computational techniques is the limited simulation sizes and timescales that can be used. Empirical potentials do away with the electrons and model the total energy as a sum of local atomic energies which depend only on the local environment of each atom. These simplifications lead to potentials that are typically many orders of magnitude faster and, crucially, a computational cost that scales linearly with the number of atoms. 
Moving from hand-crafted  potentials with tens of parameters to machine learned potentials \cite{, behler2007generalized, bartok2010gaussian} has lead to a substantial increase in the accuracy and transferability of such potentials in recent years.  \cite{bartok2018machine, zuo2020performance, shapeev2016moment, van2020regularised, faber2018alchemical}. This improvement has enabled vast, and accurate, simulations involving hundreds of thousands of atoms \cite{deringer2021origins} that would otherwise be completely inaccessible.

Another area which has seen rapid development is the application of both supervised and unsupervised learning techniques to large scale materials data \cite{jain2016research, saal2013materials, draxl2018nomad}. Such techniques have already seen great success and will undoubtedly prove useful in expediting material design and discovery \cite{bernstein2019novo} as well as understanding observed trends. Both potential fitting and, more broadly, materials machine learning require mathematical descriptions of material structures that can be used as inputs to models. Many different types of descriptors have been proposed \cite{musil2021physics}, almost all of which are invariant to translations, rotations, and permutation of equivalent atoms. Incorporating these symmetries within the descriptor avoids models having to ``learn'' them, leading to greater data efficiency. Another commonality is the rapid increase in descriptor size for environments composed of multiple elements. For instance, the size of the \ac{SOAP} power spectrum \cite{bartok2013representing} scales quadratically with the number of elements $S$ whilst the length of the bispectrum scales as $S^3$. This increase poses challenges for interatomic potentials, both in terms of evaluation speed and the memory required for fitting, as well as more generally for the storage of descriptors in databases.  Before introducing some of the existing approaches it is worth noting that the problem can be somewhat mitigated by exploiting sparsity where possible. For instance, the SOAP power spectrum is sparse with respect to elements, so that even if there are $S_{\text{total}}$ elements present across a given dataset, only those present in a given environment $S_{\text{env}}$ need be considered when computing an individual descriptor. Whilst this can facilitate a storage saving, it does not always lead to a reduction in the number of model parameters, e.g. in a linear model, nor does it fully solve the problem for higher-body order descriptors; the corresponding \ac{SOAP} bispectrum for a high entropy alloy liquid environment with $S_{\text{env}}=8$ is $\sim$500 times larger than if a single element were present.  

A great deal of effort has already gone into ways of reducing this scaling, with many interesting approaches being used. In refs. \cite{artrith2017efficient} and \cite{uhrin2021through} constant complexity in $S$ is achieved by concatenating two descriptors, one of which is element agnostic and another where the contributions from each element are weighted. This effectively amounts to embedding the elemental information into two dimensions, rather than keeping different elements entirely distinct. A similar strategy was used in refs. \cite{schutt2018schnet} and \cite{willatt2018feature}, except here the element embeddings were optimised during model fitting, so that the final embeddings contained a data-driven measure of chemical similarity between the elements. Approaching the problem from an information content perspective, the recent work of ref. \cite{glielmo2021ranking}, demonstrated that a model fitted to as few as 10\% of the power spectrum components led to negligible degradation in force errors on an independent test set, suggesting that significant compression can be achieved. Similar results were seen in ref. \cite{onat2020sensitivity}, where descriptors were selected using CUR matrix decomposition \cite{mahoney2009cur}, in ref. \cite{nigam2020recursive}, where state-of-the-art model performance was achieved on the QM9 dataset with a heavily compressed descriptor obtained though repeated tensor products and reduction using \ac{PCA}, and also in ref. \cite{goscinski2021optimal} where a data-driven approach to constructing an optimal radial basis was employed with great success. 

A complimentary strategy is to focus on general, non-data driven, ways of reducing the scaling with $S$ whilst minimizing the loss of information. Such compression strategies could prove useful in situations where the dataset evolves over time, e.g. adding configurations with new elements, or for the storage of descriptors in databases such as NOMAD \cite{draxl2018nomad} \cite{draxl2019nomad}, where the use-case for the descriptors is not yet known. Analytic compression strategies could also be used before applying data-driven compressions, allowing for larger and more diverse datasets to be treated in this way. The recently introduced \ac{MILAD} descriptors \cite{uhrin2021through} follow this philosophy by  using a minimal set of rotational invariants as the descriptor, such that the environment can be recovered by inverting the descriptor. Whilst the primary focus of that work was not to reduce the scaling with $S$, we stress that the core ideas used are highly pertinent and that similar themes are present here.

In this work we introduce two novel, non-data driven, approaches for compressing the SOAP power spectrum; the power spectrum is available through the quippy \cite{quippy} \cite{Kermode2020-wu}, dscribe \cite{dscribe} and librascal \cite{musil2021efficient} packages. Firstly, by considering the ability to recover the density expansion coefficients from the power spectrum, we introduce a compressed power spectrum and show that, under certain conditions, it is possible to recover the original descriptor from the compressed version. Secondly, we introduce a generalisation of the \ac{SOAP} kernel which affords compression with respect to both $S$ and the number of radial basis functions $N$ used in the density expansion. This kernel retains a useful physical interpretation and the ideas used are applicable to all body-ordered descriptors, which is demonstrated using the \acp{ACSF} \cite{behler2011atom}. Finally, we evaluate the performance of the compressed descriptors across a variety of datasets using numerical tests which probe the information content, sensitivity to small perturbations and the accuracy of fitted energy models and force-fields. 

\section{Results}

\subsection{SOAP}
The \ac{SOAP} kernel, introduced in ref. \cite{bartok2013representing}, provides a way of computing the similarity between a pair of atomic environments. Invariance to permutation of equivalent atoms is achieved by forming densities,

\begin{equation}
\rho^\alpha(\mathbf{r}) = \sum_i \delta_{\alpha s_i } \exp{\left[\frac{-|\mathbf{r}-\mathbf{r}_i|^2}{2\sigma^2}\right]} f_{\text{cut}}(\mathbf{r})
\end{equation}

\noindent where a separate density is constructed for each element $\alpha$, $\sigma$ controls the width of the Gaussians used to construct the density, $f_{\text{cut}}(\mathbf{r})$ is a cutoff function that ensures atoms enter the environment smoothly  and $\mathbf{r}_i$ and $s_i$ denote the the position and element of the $i$th neighbour atom respectively.  The kernel is made invariant to the orientation of environments by integrating over all possible rotations $\hat{R} \in SO(3)$ with the full multi-species kernel \cite{de2016comparing} defined as

\begin{equation}
    k(\rho, \rho') = \int d\hat{R} \left | \sum_\alpha \int d\mathbf{r} \rho^\alpha(\mathbf{r}) \rho'^\alpha(\hat{R}\mathbf{r})\right |^\nu.
    \label{eqn:full_kernel}
\end{equation}

Expanding the density using the spherical harmonics $Y_{lm}(\mathbf{\hat{r}})$, and a set of orthogonal radial basis functions $g_n(r)$, 
$$\rho^\alpha(\mathbf{r}) = \sum_{nlm} c^\alpha_{nlm} Y_{lm}(\mathbf{\hat{r}}) g_n(r)$$

\noindent and substituting into the definition of the \ac{SOAP} kernel with $\nu=2$ yields,
\begin{equation}
    \begin{split}
        k(\rho, \rho') &= \sum_{\alpha\beta}\sum_{nn'l} p^{\alpha\beta}_{nn'l} p'^{\alpha\beta}_{nn'l} = \mathbf{p} \cdot \mathbf{p'} \\
        p^{\alpha\beta}_{nn'l} &= \sum_m c^{\alpha*}_{nlm} c^{\beta}_{n'lm} =  \mathbf{c}_{nl}^{\alpha*} \cdot \mathbf{c}^\beta_{n'l}
    \end{split}
    \label{eqn:MS_powerspec}
\end{equation}
where $\mathbf{p}$ and $\mathbf{p'}$ are the \ac{SOAP} power spectrums for the environments. In this form the rotational invariance of the power spectrum can be seen by noting that density expansion coefficients transform under rotations as
$$\mathbf{c}^\alpha_{nl} \rightarrow  \mathbf{D}^l (\hat{R}) \mathbf{c}^\alpha_{nl} $$
where $D$ is a unitary Wigner-D matrix, so that the terms $ \mathbf{c}_{nl}^{\alpha*} \cdot \mathbf{c}^\beta_{n'l}$ are individually invariant \cite{bartok2013representing}. Taking $\sigma \rightarrow 0$ (for convienience) in Equation \ref{eqn:full_kernel} with $\nu=2$ reveals that the power spectrum is a 3-body descriptor,
\begin{equation}
      k(\rho, \rho') = \int d\hat{R} \sum_{ijpq} \delta_{s_i s_p}\delta(\mathbf{r}_i-\hat{R}\mathbf{r}_p ) \delta_{s_j s_q}\delta(\mathbf{r}_j-\hat{R}\mathbf{r}_q)
\end{equation}
where $i,j$ and $p,q$ index atoms in the first and second environment respectively. The integration over all rotations ensures that each matching pair of triangles, where one vertex is the central atom and the others are neighbour atoms, between the environments contributes to the kernel, so that the power spectrum is effectively a histogram of triangles. Increasing $\nu$ increases the body order of the descriptor, so that the bispectrum ($\nu=3$) corresponds to a histogram of tetrahedra and so on. With multiple elements, the neighbour atoms at the corner of each triangle must also match, so that there is a histogram of triangles for each pair of elements. In general, the length of the descriptor scales as $S^\nu$ where $S$ is the number of elements and the body-order is $\nu+1$, rendering these descriptors impractical for large $S$. In this work we investigate a number of alternatives which aim to circumvent this scaling, with a particular focus on compressing the power spectrum.
\subsection{Information Content}\label{sec:info_content}
After exploiting symmetry, $p^{\alpha \beta}_{nn'l} = p^{\beta \alpha}_{n'nl}$, the length of the power spectrum is $\frac{1}{2}NS(NS+1)(L+1)$, where $N$, $L$ and $S$ are the number of radial basis functions, highest order of spherical harmonic and total number of elements respectively. Here we show that the power spectrum is amenable to lossless compression, with the final descriptor having length $NS(L+1)^2$. We start by highlighting that the sum over $m$ in Equation \ref{eqn:MS_powerspec} can instead be viewed as a dot product between density expansions coefficient vectors $\mathbf{c}^\alpha_{nl}$ of length $2l+1$.  Then, because all products between coefficient vectors with equal $l$ index are taken, the power spectrum can be re-shaped from a vector into a sequence of ``$l$-slices''
\begin{equation*}
    P_l=\begin{pmatrix}
    \mathbf{c}^\alpha_1\cdot \mathbf{c}^\alpha_1 & \mathbf{c}^\alpha_1\cdot \mathbf{c}^\alpha_2 & ... & \mathbf{c}^\alpha_1\cdot \mathbf{c}^S_N \\
     \mathbf{c}^\alpha_2\cdot \mathbf{c}^\alpha_2 & \mathbf{c}^\alpha_2\cdot \mathbf{c}^\alpha_2 & ... & \mathbf{c}^\alpha_2\cdot \mathbf{c}^S_N \\
    \vdots& \vdots & \ddots & \vdots \\
     \mathbf{c}^\beta_1\cdot \mathbf{c}^\alpha_1 & \mathbf{c}^\beta_1\cdot \mathbf{c}^\alpha_2 & ... & \mathbf{c}^\beta_1\cdot \mathbf{c}^S_N \\
    \vdots& \vdots & \ddots & \vdots \\
    \mathbf{c}_N^S\cdot \mathbf{c}_1^\alpha & \mathbf{c}_N^S\cdot \mathbf{c}_2^\alpha & ... & \mathbf{c}_N^S\cdot \mathbf{c}_N^S\\
    \end{pmatrix}
\end{equation*}
where the $l$ index has been suppressed as it is the same for all vectors in a given slice. Viewed as such, $P_l$ is a Gram matrix between the coefficient vectors, $P_l = X_l^T X_l$ where $X_l = (\mathbf{c}^\alpha_1, \mathbf{c}^\alpha_2, \dots, \mathbf{c}^S_N)$.  As $P_l$ only contains information on the relative orientations of the $\mathbf{c}_n^\alpha$, the ${\mathbf{c}_n^\alpha}$ can, at best, be recovered from $P_l$ up to a global rotation in the $2l+1$ dimensional space. A direct consequence of this is that simultaneously rotating all coefficient vectors with equal $l$ index does not affect the power spectrum, so that there are many different densities which share the same power spectrum. Fortunately, the space of atomic densities, i.e. densities which correspond to physically plausible atomic configurations, is much smaller than the space of all densities, so that this effect is somewhat mitigated in practice.
\newline\newline Interestingly, whilst $P_l$ is an $NS\times NS$ matrix, its rank is at most $2l+1$, as the rank of the $(2l+1)\times NS$ matrix $X_l$ is at most $2l+1$. This means that provided the first $2l+1$ columns in $X_l$ form a basis, then all terms in $P_l$ can be recovered from only the first $2l+1$ rows. This observation suggests that the power spectrum is amenable to non-data-driven lossless compression when $2l+1 < NS$. For situations with many different elements $NS \gg 2l+1$ so that the potential compression factor is large. Rather than simply storing the first $2l+1$ rows, we propose storing $2l+1$ random linear combinations of all rows,

\begin{equation}
    W^TP_l=\begin{pmatrix}
    \mathbf{b}_1\cdot \mathbf{c}_1 & \mathbf{b}_1\cdot \mathbf{c}_2 & ... & \mathbf{b}_1\cdot \mathbf{c}_{NS} \\
    \mathbf{b}_2\cdot \mathbf{c}_1 & \mathbf{b}_2\cdot \mathbf{c}_2 & ... & \mathbf{b}_2\cdot \mathbf{c}_{NS}\\
    \vdots& \vdots & \ddots & \vdots \\
    \mathbf{b}_{2l+1}\cdot \mathbf{c}_1 & \mathbf{b}_{2l+1}\cdot \mathbf{c}_2 & ... & \mathbf{b}_{2l+1}\cdot \mathbf{c}_{NS}\\
    \end{pmatrix}
    \label{eqn:random_weights}
\end{equation}

where $W$ is an $NS\times(2l+1)$ matrix of randomly chosen weights.

A set of density expansion coefficients consistent with the original power spectrum can then be recovered from $W^TP_l$ by diagonalising $ W^T P_l W = U^T\Lambda U $, taking $(\mathbf{b}_1, \mathbf{b}_2, \dots, \mathbf{b}_{2l+1}) = \Lambda^\frac{1}{2}U$ and then forming $X_l = \Lambda^{-\frac{1}{2}}UW^TP_l$, where $U$ is a unitary matrix whose columns are the eigenvectors of  $W^T P_l W$ and $\Lambda$ is a diagonal matrix of the eigenvalues \cite{riley1999mathematical}. This procedure fails if $(\mathbf{b}_1, \mathbf{b}_2, \dots, \mathbf{b}_{2l+1})$ does not form a basis, so that $\Lambda$ has one or more zero eigenvalues. This could arise because $\text{rank}(W^TX_l) < \text{rank}(X_l) \le 2l+1$, which is highly unlikely because of the random weights, or because $\text{rank}(W^TX_l) =\text{rank}(X_l) = r < 2l+1$, which occurs frequently as explained below. From a recovery perspective the latter is not problematic as the same procedure can be carried out using $(\mathbf{b}_1, \mathbf{b}_2, \dots, \mathbf{b}_r)$ as a basis by discarding rows and columns from $W^ TP_l W$ as required. 

\begin{figure}[h!]
  \ifthenelse{\boolean{show_figs}}
  {\includegraphics[width=0.45\textwidth]{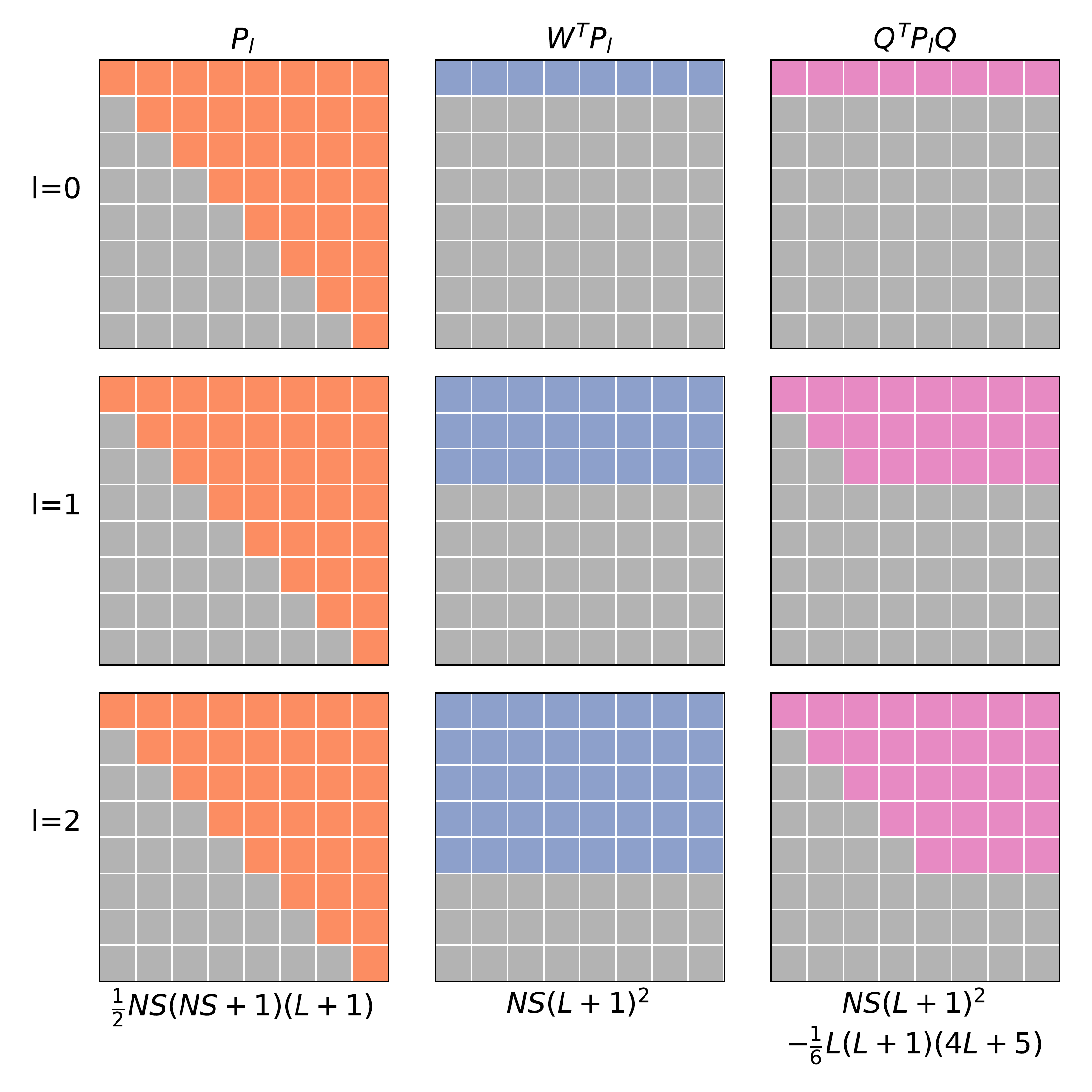}}
  {\includegraphics[width=0.5\textwidth]{example-image-a}}
  \caption{Moving from left to right $P_l$, $W^T P_l$ and $Q^TP_lQ$ are shown for $l=0,1,2$ and $NS=8$. Typically $NS$ will be significantly larger than shown here. Only the elements shown in color need be stored to determine all remaining elements which are shown in grey. The total length of each descriptor is listed underneath.}
  \label{fig:compression_saving}
\end{figure}

By compressing all ``$l$-slices'' in this way only $NS(L+1)^2$ invariants need be stored, compared to $\frac{1}{2}NS(NS+1)(L+1)$ for the uncompressed power spectrum. Interestingly, slightly more compression can be achieved by forming the symmetric matrix $Q^TP_lQ$, where $Q$ is an $NS \times NS$ matrix of random weights and then storing only the upper right triangular portion --- the original power spectrum can still be recovered in an entirely analogous fashion. The symmetry of $Q^TP_lQ$ means that there are actually fewer invariants then density expansion coefficients; per $l$ slice, there are $l(2l+1)$ fewer invariants, which corresponds exactly to the number of distinct rotations in $2l+1$ dimensions. However, the cost of this additional compression is a loss of sparsity. Whereas only the entries of $W^TP_l$ (and $P_l$) corresponding to elements present in the local atomic environment will be non-zero, $Q^TP_lQ$ will be a dense matrix as all the $\mathbf{c}^\alpha_{nl}$ are ``mixed'' together. Retaining this sparsity is exceptionally important for efficient descriptor storage as often $S \gg S_{\text{env}}$, where $S$ is the total number of elements present in a dataset whereas $S_{\text{env}}$ is the typical number present in any given environment.

Finally, we note that there is an additional restriction on the rank of $P_l$, namely $\text{rank}(P_l) \leq n_{\text{neighb}} $ where $n_{\text{neighb}}$ is the number of neighbouring atoms contained within the cutoff. This occurs because the density expansion vectors for the density of a \textit{single neighbour atom} corresponding to different radial basis functions are all parallel. This is shown explicitly in the \hyperlink{SI}{Supporting Information} but can be understood by noting that the projections of a Gaussian onto different radial shells will vary only in magnitude, so that the angular part of the expansion, which determines the direction of $\mathbf{c}^\alpha_{nl}$, will be identical. This observation is consistent with intuition -- the information content should depend on the number of neighbours -- and may prove useful when constructing concise descriptors for datasets where $n_{\text{neighb}}$ is known to be bounded.

\subsection{Generalized Kernel}
In the previous section we showed that it is possible to compress the SOAP power spectrum in a lossless manner. Here we introduce a family of physically interpretable compression options based on generalising the SOAP kernel. The first step is to generalise Equation \ref{eqn:full_kernel} to
\begin{equation}
     k(\rho, \rho') = \int d\hat{R}  \left | \sum_\alpha \int d\mathbf{r} \rho^\alpha(\mathbf{r}) \rho'^\alpha(\hat{R}\mathbf{r})\right|^\nu \left | \int d\mathbf{r} \rho(\mathbf{r}) \rho'(\hat{R}\mathbf{r})\right |^\mu 
     \label{eqn:general_kernel}
\end{equation}
where the first factor is as before and the second factor involves the total density $\rho(\mathbf{r}) = \sum_\alpha \rho(\mathbf{r})^\alpha$. The advantage of this approach is that the total body order is now partitioned into element-sensitive and element-agnostic terms, so that the scaling with $S$ can be controlled separately from the overall body-order. For instance, $\nu,\mu=1,1$ results in a modified power spectrum $p^\alpha_{nn'l}$ with a length that scales linearly with $S$ and is related to the original power spectrum, $\nu=2$, via,
\begin{equation}
    p^\alpha_{nn'l} = \sum_{m}c^{\alpha*}_{nlm} \left( \sum_\beta c^\beta_{n'lm}\right) = \sum_\beta p^{\alpha\beta}_{nn'l}
    \label{eqn:nu=1mu=1}
\end{equation}

\begin{figure*}[htp]
  \includegraphics[width=0.7\textwidth]{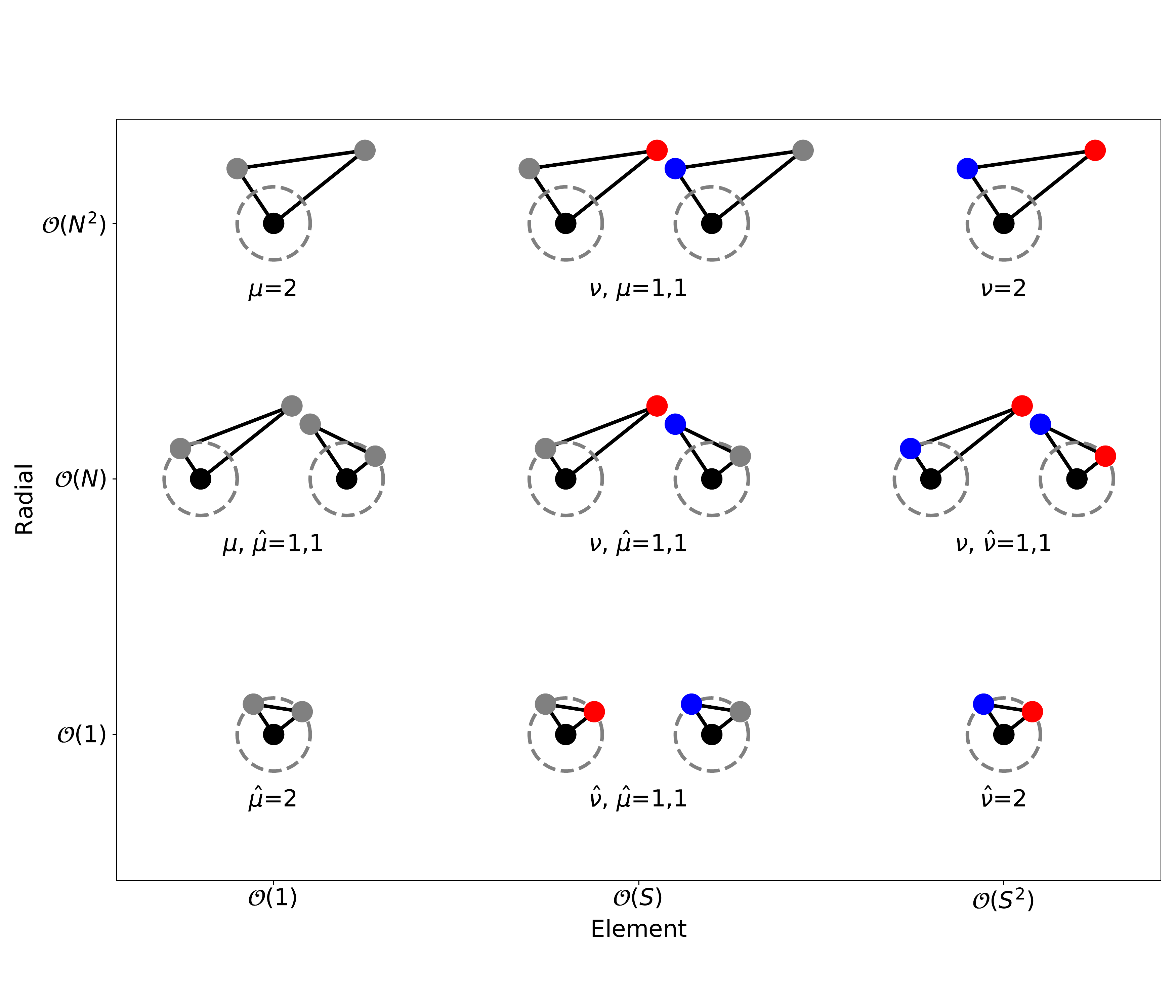}
  \caption{Schematic showing the physical interpretation of the generalised power spectrum for various choices of $\nu$, $\hat{\nu}$, $\mu$ and $\hat{\mu}$. For such 3-body terms $\nu + \hat{\nu} + \mu + \hat{\mu} = 2$; indices which are zero are not listed. The vertices shown in blue and red are element-sensitive whilst those shown in grey are not. The grey dashed line indicates the unit sphere. When the projection results in two distinct triangles both are shown, otherwise only one is shown.}
  \label{fig:3-body}
\end{figure*}

The  $\nu,\mu=1,1$ power spectrum still corresponds to a histogram of triangles, but now only one vertex of each triangle is element-sensitive, as shown in Figure \ref{fig:3-body}. Clearly, this idea can be applied just as well for higher body-orders to provide compression with respect to $S$. In the previous section we also achieved compression with respect to $N$. Following a similar approach as for $S$, we further generalise the kernel to

\begin{equation}
\begin{split}
    & k(\rho, \rho') =  \int d\hat{R}  \left| \int d\mathbf{r} \rho(\mathbf{r}) \rho'(\hat{R}\mathbf{r})\right |^\mu 
    \left| \int d\mathbf{r} \hat{\rho}(\mathbf{\hat{r}}) \hat{\rho}'(\hat{R}\mathbf{\hat{r}}) \right |^{\hat{\mu}} \\
    & \left | \sum_\alpha \int d\mathbf{r} \rho^\alpha(\mathbf{r}) \rho'^\alpha(\hat{R}\mathbf{r})\right|^\nu 
     \left | \sum_\alpha \int d\mathbf{r} \hat{\rho}^\alpha(\mathbf{\hat{r}}) \hat{\rho}'^\alpha(\hat{R}\mathbf{\hat{r}}) \right|^{\hat{\nu}}  
     \label{eqn:fully_general_kernel}
\end{split}
\end{equation}

where $\hat{\rho}(\mathbf{\hat{r}})$ is the projection of the density onto the surface of the unit sphere. As before,   Equation \ref{eqn:fully_general_kernel} still corresponds to comparing all possible triangles for $\nu+\hat{\nu}+\mu+\hat{\mu}=2$. However now some of the vertices may be projected onto the surface of the unit sphere, as well as potentially being element insensitive. The full range of 3-body compression options is depicted in Figure \ref{fig:3-body}, with the interpretation of the original power spectrum, $\nu=2$ (only non-zero index values are listed), shown in the upper right corner.

\begin{figure}[h!]
  \includegraphics[width=0.4\textwidth]{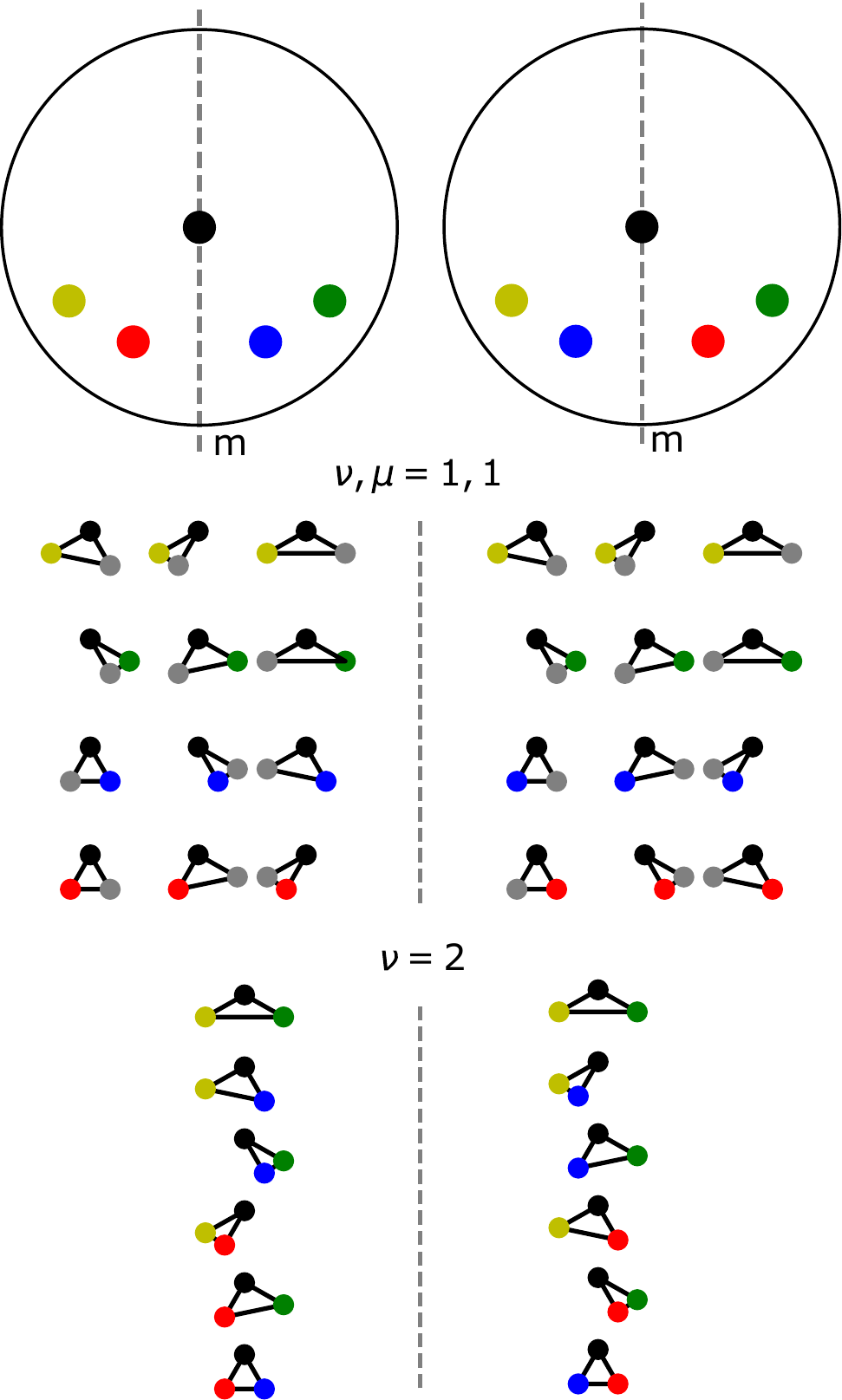}
  \caption{Two environments which are distinct using $\nu=2$ but identical according to $\nu,\mu=1,1$, because of the mirror symmetry of the total density, are shown. Elements are distinguished by color whilst the histograms of triangles shown on the right - element agnostic vertices are shown in grey.}
  \label{fig:species_degen}
\end{figure}

Utilising these compressions offers a significant reductions in descriptor size but does result in a descriptor that is typically less informative. An example of such an information loss is shown in Figure \ref{fig:species_degen}, where the two environments shown are distinguished by $\nu=2$ but not by $\mu,\nu=1,1$, as now there is only 1 element-sensitive vertex per triangle, rather than two. Returning to the idea of the previous section, we see that $\nu,\mu=1,1$ corresponds to using $\left\{\mathbf{C}_n\right\}$, where $\mathbf{C}_n = \sum_\alpha \mathbf{c}^\alpha_n$, in place of the random basis $(\mathbf{b}_1, \mathbf{b}_2, \dots, \mathbf{b}_{2l+1})$. In this case the mirror symmetry of the total density means that $\text{rank}(\mathbf{C}_1, \mathbf{C}_2, \dots, \mathbf{C}_{N}) = 2 < \text{rank}(\mathbf{c}_1, \mathbf{c}_2, \dots, \mathbf{c}_{NS}) = 4$, so that the full power spectrum cannot be recovered from the compression. However, this pair of environments are a handcrafted example and more generally, in the absence of similar symmetries, we would expect $\left\{\mathbf{C}_n\right\}$ to span the required space, provided $N \ge 2l+1$, so that the compression would be lossless. This line of thought motivates the introduction of a further compression, denoted $\nu,\mu^*=1,1$, where the kernel is as defined in Equation \ref{eqn:general_kernel}, but $2l+1$ radial basis functions are used for the total density expansion. This choice achieves the same level of compression as $W^TP_l$ and, provided $\text{rank}(\mathbf{C}_1, \mathbf{C}_2, \dots, \mathbf{C}_{2l+1}) = \text{rank}(\mathbf{c}_1, \mathbf{c}_2, \dots, \mathbf{c}_{NS})$, it will also be lossless. Such arguments suggest that compressing beyond $\nu,\mu^*=1,1$ will necessarily be lossy, with the analysis for higher body orders being left to future work.

\begin{table*}[htp]
	\centering
	\caption{Various compressions of the SOAP power spectrum are listed, $N$, $S$, $L$, $d_J$ and $u_{\text{max}}$ are the number of radial basis functions, elements, maximum order of spherical harmonics, number of embedding dimensions and the number of optimised radial basis functions respectively. The circumflex indicates projection onto the unit sphere so that $\hat{\rho}(\mathbf{r}) = \hat{\rho}(\mathbf{\hat{r}})$. The kernel for $\nu,\mu^*=1,1$ is identical as for $\nu,\mu=1,1$, except that $2l+1$ radial basis functions are used in the total density expansion, so that $n'$ = $1, 2, \dots, 2l+1$.}
	\label{table:form_complexity}
	\begin{tabular}{|c|c|c|c|}
	    \hline
	    Label & References & Descriptor & Length  \\
		\hline
		 $\nu=2$ &  Full SOAP kernel \cite{bartok2013representing, de2016comparing} &  $p^{\alpha\beta}_{nn'l} = \sum_m c^{\alpha*}_{nlm} c^{\beta}_{n'lm}$ & $\frac{1}{2}NS(NS+1)(L+1)$ \\
		 &&&\\
        $\nu,\mu=1,1$ & - &  $p^\alpha_{nn'l} = \sum_{m}c^{\alpha*}_{nlm} \left( \sum_\beta c^\beta_{n'lm}\right)$ & $N^2S(L+1)$\\
         &&&\\
          $\nu,\mu^*=1,1$ & - &  $p^\alpha_{nn'l} = \sum_{m}c^{\alpha*}_{nlm} \left( \sum_\beta c^\beta_{n'lm}\right)$ & $NS(L+1)^2$\\
         &&&\\
        $\nu,\hat{\nu}=1,1$ & \makecell{Similar to  $p^{\alpha\beta}_{1n'l} = \sum_{m}c^{\alpha*}_{1lm} c^\beta_{n'lm}$ \\ used in TurboGAP \cite{caro2019optimizing, turbogap_compress}} &  $p^\alpha_{nl} = \sum_{m}c^{\alpha*}_{nlm} \left( \sum_{n'} c^\beta_{n'lm}\right)$ & $NS(L+1)$\\
         &&&\\
        $\nu, \hat{\mu}=1,1$ & - &  $p^\alpha_{nl} = \sum_{m}c^{\alpha*}_{nlm} \left( \sum_{n'\beta } c^\beta_{n'lm}\right)$ & $NS(L+1)$\\
         &&&\\
        $\mu=2$  & Element agnostic SOAP kernel  \cite{bartok2013representing}
        &  $p_{nn'l} = \sum_{m} \left( \sum_\alpha c^{\alpha*}_{nlm} \right) \left( \sum_\beta  c^\beta_{n'lm}\right)$ & $\frac{1}{2}N(N+1)(L+1)$\\
         &&&\\
        $\nu=2$, $d_J$ &  \makecell{ Element weighting \cite{artrith2017efficient} \\ Element embedding \cite{willatt2018feature}} &  $p^{JJ'}_{nn'l} = \sum_m c^{J*}_{nlm} c^{J'}_{n'lm}$ & $\frac{1}{2}Nd_J(Nd_J+1)(L+1)$ \\
        &&&\\
        --- &  \makecell{Optimal Radial Basis \cite{goscinski2021optimal} \\ with mixed-species basis}  &  $p_{uu'l} = \sum_m c^*_{ulm} c_{u'lm}$ & $\frac{1}{2}u_{\text{max}}(u_{\text{max}}+1)(L+1)$ \\
         &&&\\
        $W^TP_l$ &  -  & $W^TP_l$ & $NS(L+1)^2$ \\
		\hline
	\end{tabular}	

\end{table*}

At first glance the degeneracies introduced by choosing $\nu,\mu=1,1$ appear catastrophic. However, it should be noted that the original power spectrum $\nu=2$ has similar degeneracies. A simple, single element example was given in ref. \cite{pozdnyakov2020incompleteness}, which is trivially modified to contain multiple elements in Figure \ref{fig:bonus_degen}. Despite this, the full $\nu=2$ power spectrum has been widely used with great success across a wide variety of applications \cite{bartok2018machine, deringer2020general, bartok2017machine}, as have other descriptors which are known to be mathematically incomplete \cite{artrith2016implementation, behler2011atom}. As such, whilst it is useful to understand the origin of any additional degeneracies in compressed descriptors, the most practically interesting question is whether or not they lead to a significant degradation in model performance across typical datasets. We assess this for the variants of the SOAP power spectrum listed in Table \ref{table:form_complexity} by performing numerical tests which are discussed in the results section. 

Finally, whilst we only test compressions of the power spectrum in this work, these compression ideas can be applied just as easily to higher body-orders. For instance, choosing $\nu, \hat{\nu},\mu=1,1,1$ would result in a compressed version of the bispectrum that scales quadratically with $S$ and $N$, rather than cubically. In general, the length of the descriptor will scale as $S^{\nu+\hat{\nu}}N^{\nu+\mu}$ with the total body order given by $\nu+\hat{\nu}+\mu+\hat{\mu}$, so that the scaling with $S$ and $N$ can be chosen independently from the overall body-order. We anticipate these compressions being particularly useful for descriptors such as \ac{ACE} \cite{drautz2019atomic} \cite{van2020regularised} where $\nu\ge4$ is commonly used.

\subsection{Element Embedding}

An alternative approach to constructing concise descriptors for systems with large $S$ was used in refs. \cite{artrith2017efficient} \cite{uhrin2021through} \cite{gastegger2018wacsf}. Rather than using a separate density for each element they instead used two density channels; the total, element agnostic, density and an element weighted density defined as $\rho^Z(\mathbf{r}) = \sum_\alpha w_\alpha \rho_\alpha(\mathbf{r})$ where $w_\alpha$ is an element dependent weight. The descriptor can then be formed using these two density channels, so that the length of the power spectrum is $N(2N+1)(L+1)$, independent of $S$.

The element-weighted power spectrum is an instance of a more general type of constant complexity approach, where each element $\alpha$ is represented by a vector $\mathbf{u}^\alpha$, where $\text{dim}(\vec{u}_\alpha)=d_J < S$, so that the chemical elements are effectively embedded into a lower dimensional space. The $\mathbf{u}^\alpha$ can then be optimised during model fitting so that the alchemical similarity between different elements, $k^{\alpha\beta}=\mathbf{u}_\alpha \cdot \mathbf{u}_\beta$ \cite{willatt2018feature}, is learned from the data. This approach was used in refs. \cite{willatt2018feature} and \cite{schutt2018schnet}, where in both cases the optimised embedding was consistent with known chemical trends, and more recently, similar, learnable mappings have been used in refs. \cite{batzner2021se}, \cite{anderson2019cormorant} and \cite{e3nn} \cite{weiler20183d} \cite{thomas2018tensor}. 

This approach was taken further still in ref. \cite{goscinski2021optimal} where \ac{PCA} was used to determine a reduced optimal radial basis for a given dataset. By allowing basis changes, followed by truncation's, that also mix different elemental channels this approach can be seen as a simultaneous embedding of both the elemental and radial information into a lower dimensional space. Interestingly, the random weight matrix $W$ in Equation \ref{eqn:random_weights} can be interpreted as performing an analogous embedding, $X_l = (\mathbf{c}^\alpha_1, \mathbf{c}^\alpha_2, \dots, \mathbf{c}^S_N) \rightarrow (\mathbf{b}_1, \mathbf{b}_2, \dots, \mathbf{b}_{2l+1}) = X_l W $. This identification connects the approaches and motivates further work where $W$ is optimised for a given dataset.

Using element embedding, with fixed or optimisable embedding vectors, offers an alternative route to reducing the scaling with $S$ and we include a corresponding compressed power spectrum, denoted as $\nu=2,d_J$, in our numerical tests for comparison. A summary of all ways of compressing the power spectrum is given in Table \ref{table:form_complexity}, with the results given in the next section. The datasets used are described briefly in section \ref{sec:datasets}.

\subsection{Distance-Distance Correlation and Information Imbalance}
\begin{figure}[h!]
  \ifthenelse{\boolean{show_figs}}
  {\includegraphics[width=0.48\textwidth]{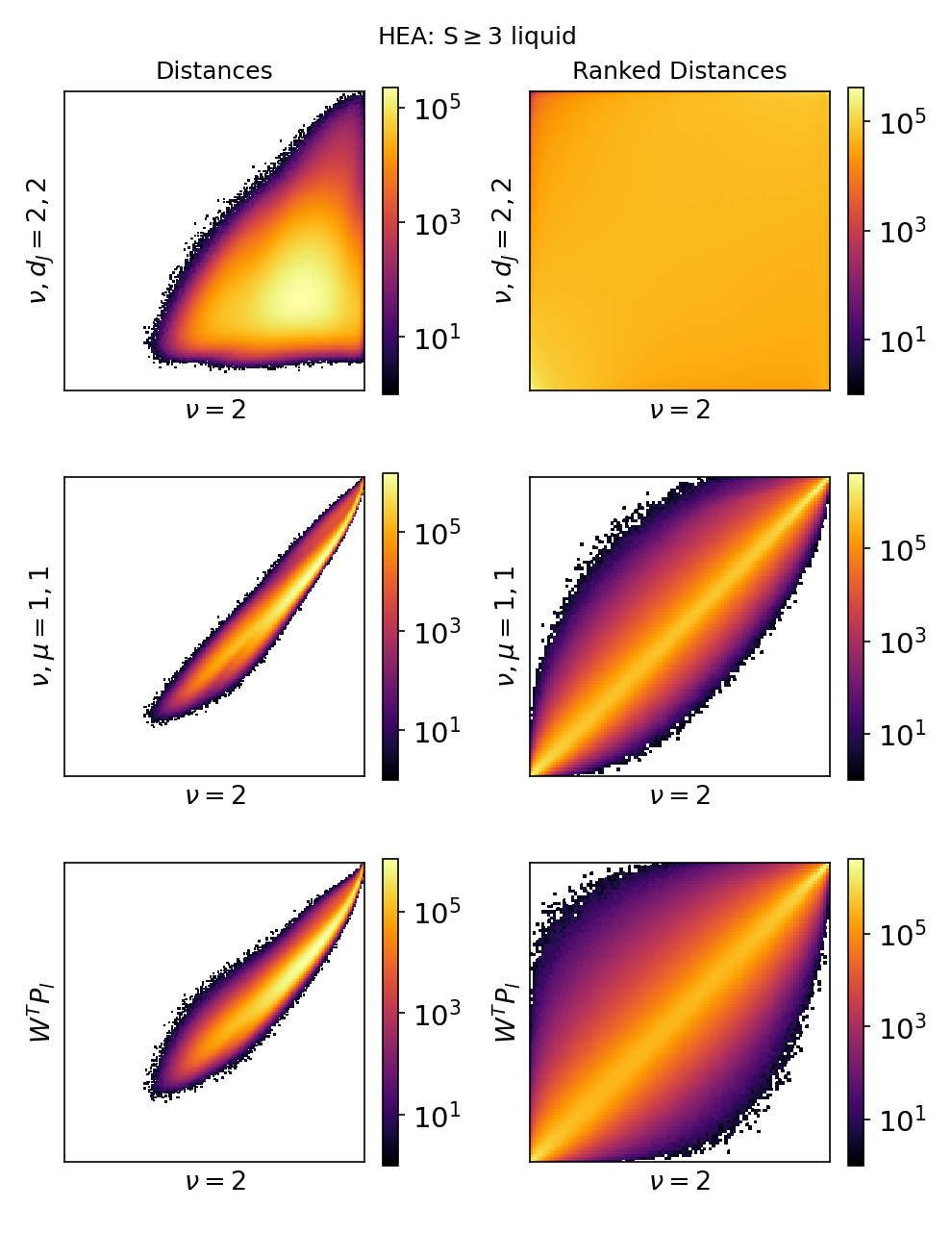}}
  {\includegraphics[width=0.95\textwidth]{example-image-a}}
  \caption{The correlation between the distances (left column) and ranked distances (right column) between all pairs of environments within liquid configuration with $S\geq3$ in the HEA dataset are shown. See Figure \ref{Sfig:HEA_correlation} for the correlations for $\mu=2$, $\nu,\hat{\mu}=1,1$ and $\nu,\mu^*=1,1$}
  \label{fig:HEA_correlation_small}
\end{figure}

The similarity between a pair of environments can be quantified by the Euclidean distance between their respective descriptor vectors, so that different descriptors give different measures of similarity. These distances can be used to quantify the relative information content of descriptors \cite{glielmo2021ranking} and, more generally, to compare how different descriptors encode a given dataset. We follow ref. \cite{parsaeifard2021assessment} using distance-distance correlation plots to compare the distances implied by the compressed descriptors to those of the full power spectrum, where it is desirable for a compression to preserve the distances as faithfully as possible. In Figure \ref{fig:HEA_correlation_small} the distance-distance (left) and ranked distance-ranked distance (right) correlations between the $\nu=2$ power spectrum and two of the compressed alternatives are shown for all pairs of environments within liquid configurations, $S\geq3$, in the HEA dataset. The advantage of comparing the ranked distances is that the ranking process eliminates scaling and monotonic transformations of the distances, leaving only the correlation structure behind \cite{glielmo2021ranking, calsaverini2009information}.

The correlations between both the distances and ranked distances for the $\nu,\mu=1,1$ compression and $\nu=2$ are strong, with a notable absence of points in the top left and bottom-right of each plot. The same is not true of both the constant complexity alternatives, where there are many environments deemed well separated by $\nu=2$ but which are poorly distinguished by $\nu,d_J=2,2$ (and $\mu=2$). An example of such an environment is shown in Figure \ref{fig:HEA_degen}, with the distances according to each descriptor listed alongside. The most striking difference is seen in the ranked-distance correlation plots which are near uniform for $\nu,d_J=2,2$ (and $\mu=2)$, demonstrating the original correlation structure is almost completely lost. 

\begin{figure}[h]
  \includegraphics[width=0.4\textwidth]{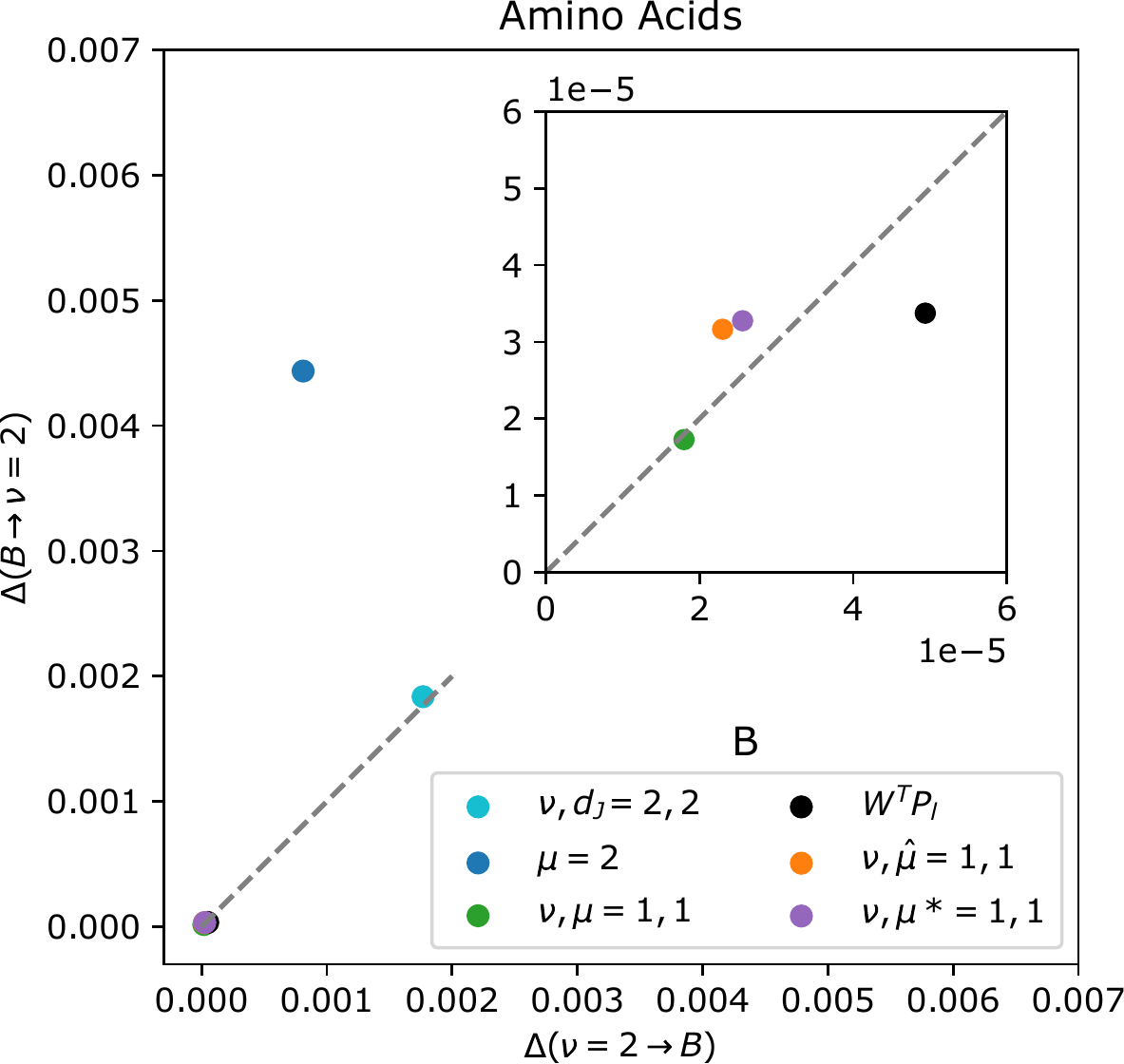}
  \caption{Information imbalance plane for atomic environments in the amino acid dataset. The central atom was not included in the density expansion.}
  \label{fig:amino_II}
\end{figure}

\begin{figure}[h!]
  \includegraphics[width=0.38\textwidth]{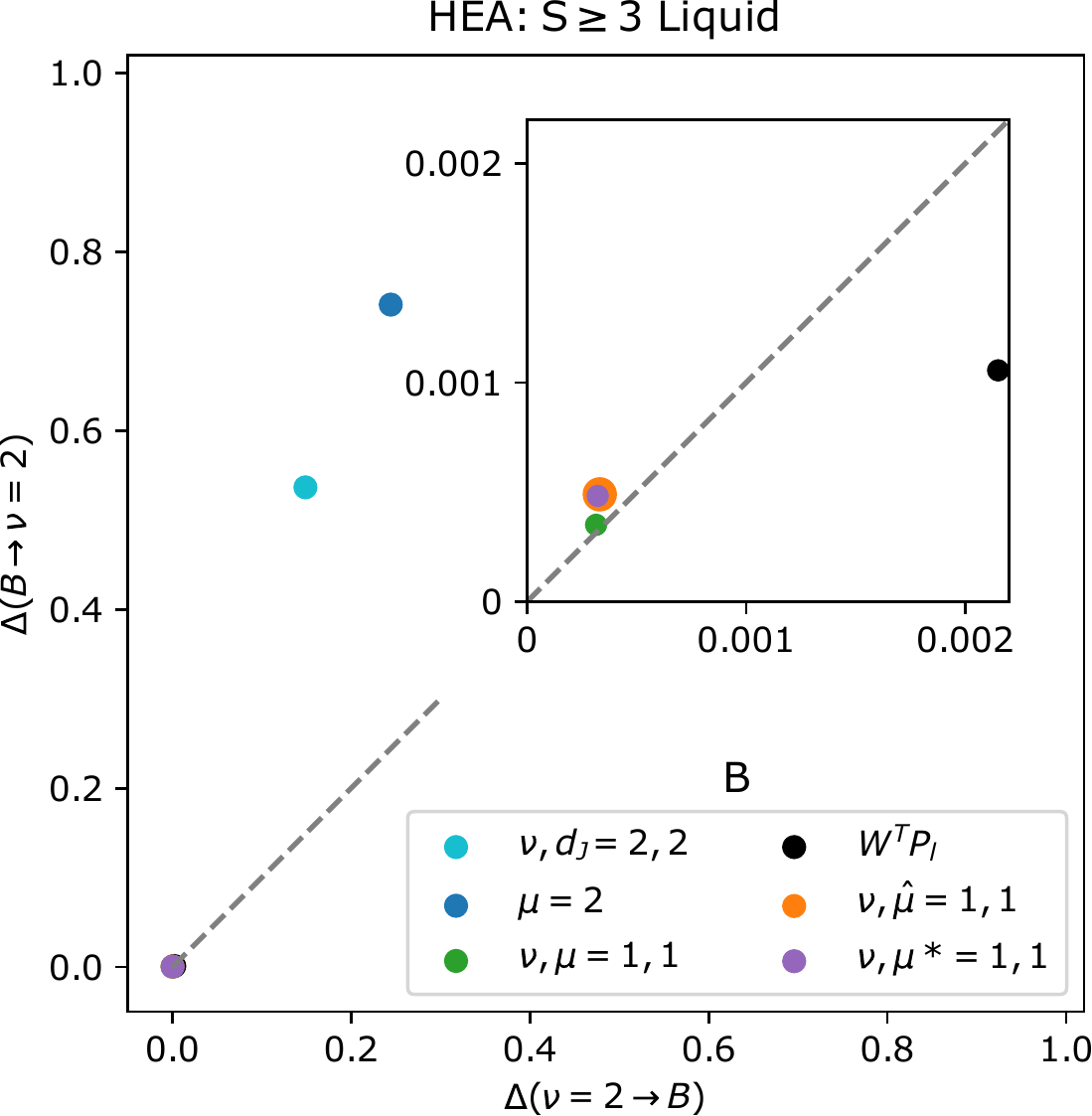}
  \caption{Information imbalance plane for the environments from liquid configurations with $S\geq3$ in the HEA dataset. Note the difference in scale relative to the amino acids.}
  \label{fig:HEA_II}
\end{figure}

We also compute the information imbalance (introduced in ref. \cite{glielmo2021ranking}) between the different descriptors; the information imbalance is a way of measuring the relative information content of different distances measures, see Section \ref{sec:methods_II} for details. The information imbalance planes for the HEA and amino acid datsets are shown in Figures \ref{fig:amino_II} and \ref{fig:HEA_II}, where points in the bottom left, top left and along the diagonal indicate descriptors which encode the same, less and orthogonal (different) information to $\nu=2$ respectively. The results for the HEA dataset provide quantitative evidence for the trends seen in Figure \ref{fig:HEA_correlation_small}, demonstrating that for this dataset, the $\nu, d_J=2,2$ (and $\mu=2)$ descriptors are significantly less informative than the others. Conversely, these descriptors carry almost identical information to $\nu=2$ on the amino acid dataset. We believe this is because all the amino acid molecules are geometry optimised, so that the atom type information can be inferred from the atomic positions alone; this is backed up by $\Delta(\mu=2 \rightarrow  \nu=2) \ll 1$. The stark differences seen between these two datasets suggests that whilst low-dimensional element embeddings are undoubtedly useful, they are not suitable as information preserving compressions for all datasets.

\subsection{Fitting to Energies}

The ultimate test of any descriptor is the accuracy and overall performance of the models which use it. In this work we test the proposed SOAP compressions by fitting interatomic potentials to total energies for four separate datasets using both \ac{RR} and \ac{KRR} models, see Section \ref{sec:models} for details.

\begin{figure*}[htp]
    \ifthenelse{\boolean{show_figs}}
    {\includegraphics[width=0.95\textwidth]{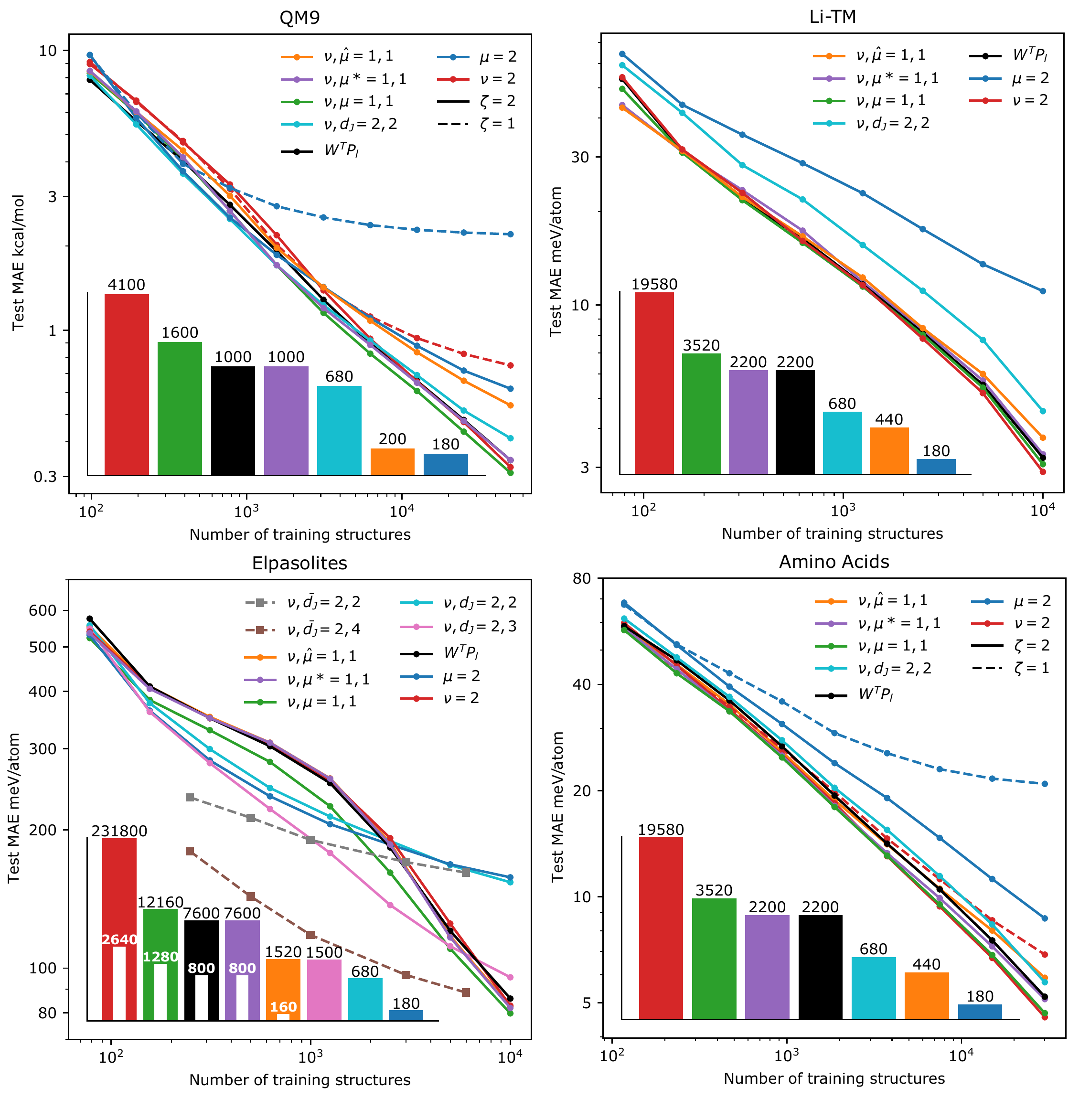}}
    {\includegraphics[width=0.95\textwidth]{example-image-a}}
  \caption{Learning curves for the total energy fits are shown. For the Li-TM dataset $\zeta=1$, 2, 4 and 8 were all tested but only the models with $\zeta=8$ are shown. The $\bar{d_J}=2$ and $\bar{d_J}=4$ curves for the Elpasolites are taken from ref. \cite{willatt2018feature} where the embedding was optimised. All models for the Elpasolites were fitted using $\zeta=1$. The total length of the descriptors is indicated in the bar chart. The overlaid white bars indicate the number of non-zero elements present, computed using $S_{\text{env}}=4$.}
  \label{fig:Energy_fits}
\end{figure*}

Learning curves for all models and datasets are shown in Figure \ref{fig:Energy_fits} whilst the total length of each descriptor is shown in the bottom left of each plot. It is important to note that all descriptors other than $\nu=2, d_J$ are sparse with respect to the elements, so that in practice their storage requirements scale with $S_{\text{env}}$, the typical number of elements present within each environment, rather than $S_{\text{total}}$, the total number of elements present in the dataset. For the elpasolite dataset $S_{\text{env}}=4$ for all structures, so in this case the lengths of the sparse descriptors are indicated in white. However, whilst the descriptors themselves are sparse the number of model parameters is dependent on the full length of the descriptor. As such, we stress that both sparsity in $S_{\text{env}}$ and reducing the scaling with respect to $S_{\text{total}}$ are highly desirable.  As a final point, the amino acids and Li-TM datasets were also used in ref. \cite{artrith2017efficient} to test the accuracy achievable combining a constant complexity descriptor and a neural network potential. We stress that the errors reported there are on the training set, acting as a proof of principle, whilst the larger errors reported here are on a distinct test set, so that the numbers are not comparable.

The trends seen across the QM9, Li-TM and amino acid datasets are broadly similar, with $\nu,\mu=1,1$ achieving the same accuracy as $\nu=2$ with a significantly reduced descriptor size. The results with $W^TP_l$ and $\nu,\mu^*=1,1$ are $\sim$5-15\% less accurate then with $\nu=2$ but, crucially, both these descriptors offer additional compression with respect to the radial channels and, as outlined in the previous section, contain sufficient information to recover the full power spectrum under certain conditions. Compressing beyond the recoverable limit with $\nu,\hat{\mu}=1,1$ --- projecting the element agnostic vertex onto the unit sphere --- offers an additional factor of $\sim$5 in terms of compression. On the elpasolite dataset this does not compromise the accuracy at all whilst for QM9 this compression incurs a 66\% increase in the MAE, which is limited to $\sim$30\% for both the Li-TM's and amino acids. Comparing these results to the errors achieved using $\mu=2$  and $\nu=2$ hints at the relative importance of atom type and geometric information for the different datasets. Unsurprisingly, knowledge of the atom types is much more important for the elapsolites, where all structures are in the same crystal prototype, whilst for the chemically reasonable organic molecules found in the QM9 and the amino acid datasets a reasonable model can be fit using geometric information alone - MAE of 0.62 kcal/mol for $\mu=2$ on QM9. This is consistent with the information imbalance analysis on the amino acids and suggests that for these molecules, in their equilibrium geometries, it is possible to use known bond lengths and coordination numbers to infer the atom type information from the geometry alone.

The (unoptimised) embedding $\nu,d_J=2,2$ performs relatively well with errors 27\%, 57\% and 26\% larger then $\nu=2$ for the QM9, Li-TM, and amino acids datsets respectively. This performance is most comparable to $\nu,\hat{\mu}=1,1$, which, particularly after exploiting sparsity, offers a greater level of compression. A clear, qualitative difference in behaviour is seen for the embedding approaches on the elpasolite dataset, which contains a much larger number of elements. Here, the unoptimised embedding performs no better than using $\mu=2$ and has a greatly diminished final learning rate compared to $\nu=2$. For comparison, the optimised embeddings, denoted using a $\bar{d_J}$, from ref. \cite{willatt2018feature}, which use similar SOAP parameters ($r_{\text{cut}}=5$\AA, $N=12$, $L=9$), are overlaid. This shows that optimising the embedding with only two dimensions leads to minimal improvement although a significant gain can be made using $\bar{d_J}=4$. However, the final learning rate for $\bar{d_J}=4$ is still diminished, relative to $\nu=2$, again indicating a qualitative difference in behaviour from the compressions which keep separate densities for each element. It is also worth noting that as $S_{\text{env}}=4$ for the elpasolites, using $d_J=4$ does not offer any descriptor compression compared to $\nu=2$ once sparsity is exploited. The embedding approaches do however offer a clear advantage in the low-data regime, which we believe is due to the reduced dimensionality of the descriptor space. If each data point occupies a certain volume of descriptor space then such embeddings greatly increase the relative fraction of ``physically reasonable'' descriptor space which is occupied by each point. As such, the relevant descriptor space is ``covered'' faster so that shortest distance from a test data point to any point in the training set decreases faster with new training data.

\subsection{Fitting a Force-field}
\begin{figure}[h!]
    \ifthenelse{\boolean{show_figs}}
    {\includegraphics[ width=0.48\textwidth]{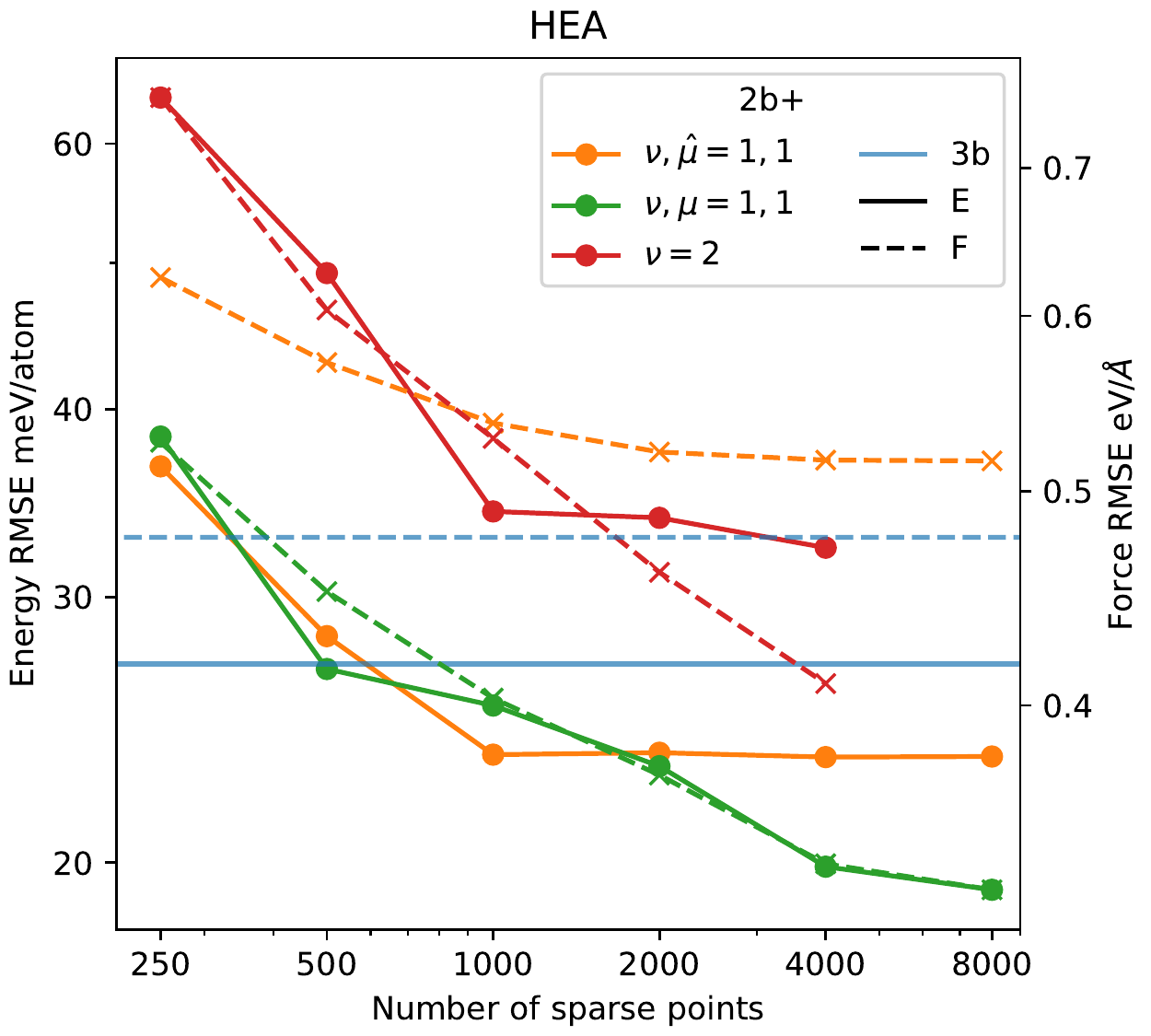}}
    {\includegraphics[width=0.45\textwidth]{example-image-a}}
  \caption{ The left hand panel shows how the energy and force errors on the test set change with the number of sparse points. Note that the 2b+3b model is from ref. \cite{byggmastar2021modeling} where it was reported that increasing the number of sparse points did not improve performance.}
  \label{fig:HEA_fits}
\end{figure}

The compression options illustrated in Figure \ref{fig:3-body} were tested further by training sparse KRR models on energies, forces and stresses for the HEA dataset using the gap\_fit program \cite{gapfit}, with the same parameters used in ref. \cite{byggmastar2021modeling}. This dataset is of particular interest as in previous work, ref. \cite{byggmastar2021modeling}, it was found that using a model fitted using a 2-body + 3-body  descriptor performed better than using a 2-body descriptor + the full multi-element power spectrum, $\nu=2$. This result is explained by the power spectrum being a higher dimensional descriptor, so that much more training data is required to span the full descriptor space corresponding to all physically relevant atomic environments. In the low data regime, i.e. before this condition is satisfied, the model will generalise very poorly outside of the training set. However, provided sufficient data, one would expect a model trained on a more informative descriptor to provide greater accuracy. This effect can be seen in the left hand panel of Figure \ref{fig:HEA_fits} where the force errors for $\nu,\hat{\mu}=1,1$ are initially better than for $\nu=2$, but then plateau quickly. In contrast the errors for the $\nu=2$ model continue to decrease, however, here the very large descriptor meant that only 4000 sparse points could be used due to a memory restriction of 1.5 TB of RAM. 

Interestingly the best model, subject to the practical restrictions on memory and quantity of training data, makes use of the $\nu,\mu=1,1$ compression. Whilst the increase in performance, compared to the 2b+3b, is modest across the entire test set it is much more dramatic on the quinary alloy liquid configurations where the energy RMSE is reduced from 96 meV/atom to 12.3 meV/atom, Figure \ref{fig:HEA_errors}.

\subsection{Sensitivity Analysis }
When using atomic descriptors for tasks such as regression it is highly desirable that they are sensitive to small perturbations of a given atomic environment, with in depth analysis of well known descriptors carried out in refs. \cite{onat2020sensitivity} and \cite{parsaeifard2021assessment}. The latter introduced the concept of the sensitivity matrix $\Lambda$, see section \ref{sec:methods_sensitivity}, where the eigenvalues $\lambda$ can be used to determine if there are any perturbations of the local environment which a descriptor is insensitive to.

\begin{figure}[h!]
    \ifthenelse{\boolean{show_figs}}
    {\includegraphics[width=0.45\textwidth]{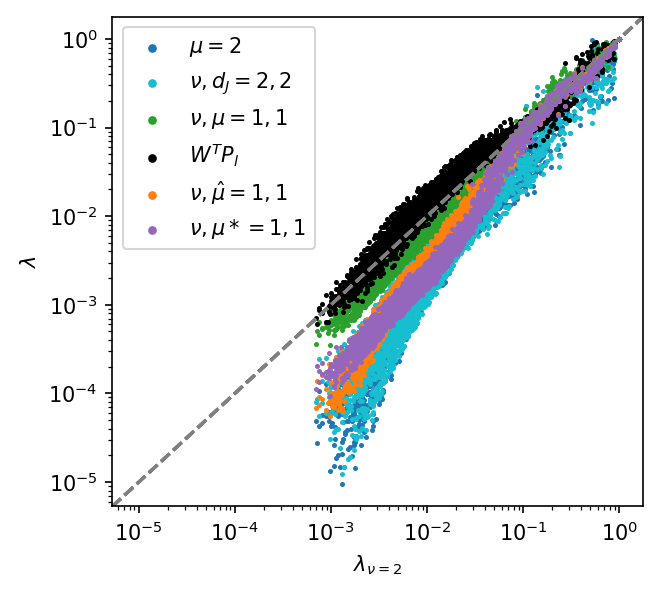}}
    {\includegraphics[width=0.5\textwidth]{example-image-a}}
  \caption{A scatter plot showing eigenvalues, $\lambda$, of the sensitivity matrix for 50 randomly chosen liquid environments from the HEA dataset. The first 6 eigenvalues are not shown. The grey dashed line indicates $y=x$. The eigenvalues have been scaled, so that the largest eigenvalue for each environment is 1. }
  \label{fig:sensitivity_correlation}
\end{figure}

The results in Figures \ref{fig:sensitivity_correlation} and \ref{fig:sensitivity_overlay} show that the proposed compressions do not significantly affect the sensitivity. In particular, no perturbations were found for which any of the compressed descriptors had near-zero sensitivity, whilst the original $\nu=2$ did not. This behaviour is not entirely unexpected, as for instance, choosing $\mu=2$ is equivalent to using the original power spectrum with a single element environment. Perhaps more surprising is that using $\nu,\hat{\mu}=1,1$ --- projecting one atom onto the unit sphere --- did not reduce the sensitivity more drastically. Of course, these tests are not exhaustive and it is probable that special environments, likely closely related to any additional degeneracies, exist where this is not the case \cite{pozdnyakov2021local}. However, we find these results promising and leave further investigation to future work.

\subsection{Atom Centered Symmetry Functions}
Whilst the compression presented in Section \ref{sec:info_content} is specific to the \ac{SOAP} power spectrum, the ideas used to form the generalized SOAP kernel are equally applicable to all body-ordered descriptors. To demonstrate this we fit KRR models to the QM9 and Li-TM datasets using compressed versions of the popular \acp{ACSF} introduced in ref. \cite{behler2011atom}. We also fit models based on element agnostic and element-weighted \acp{ACSF} \cite{artrith2017efficient} \cite{gastegger2018wacsf} for comparison. As before a polynomial kernel, $k(\boldsymbol{x}_i, \mathbf{x}_j) = \left(\mathbf{x}_i \cdot \mathbf{x}_j \right)^\zeta $, was employed and the regularisation strength was chosen using k-fold cross-validation with k=10. The G2 (2-body) and G4 (3-body) \ac{ACSF} functions were used with the traditional parameters \cite{behler2011atom, onat2020sensitivity} scaled to cutoffs of 4\AA\space and 6\AA\space for the QM9 and Li-TM datasets respectively. Compression was applied to the 3-body terms only and the descriptors are labelled using the notation outlined in Figure \ref{fig:3-body}, so that $\nu=2$ is the conventional $\vec{\text{G4}}^{\alpha\beta}$ which has length $\mathcal{O}(S^2)$, $\nu,\mu=1,1$ corresponds to $\vec{\text{G4}}^{\alpha} = \sum_\beta \vec{\text{G4}}^{\alpha\beta}$ so has length $\mathcal{O}(S)$ and $\mu=2$ is element agnostic, $\vec{\text{G4}} = \sum_{\alpha\beta} \vec{\text{G4}}^{\alpha\beta}$. For consistency with the existing literature we label the element-weighted \ac{ACSF} as wACSF \cite{gastegger2018wacsf}, rather than $\nu,d_J=2,2$. The dscribe \cite{dscribe} python package was used to compute the full \ac{ACSF} descriptor, $\nu=2$, and the compressed variants were formed by summing over elements as required.

\begin{figure*}[htp]
   \begin{minipage}{0.48\textwidth}
        \ifthenelse{\boolean{show_figs}}
        {\includegraphics[width=0.99\textwidth]{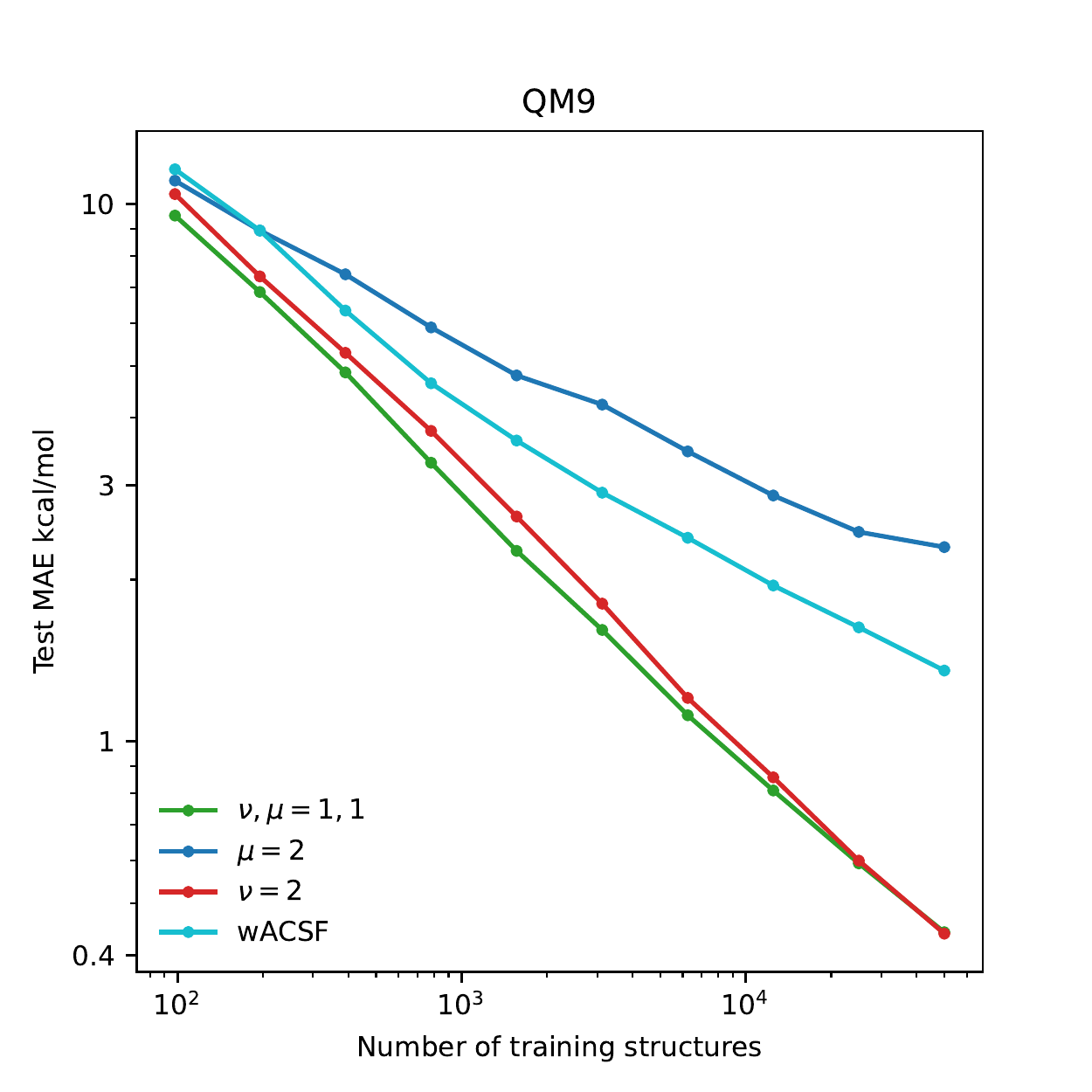}}
        {\includegraphics[width=0.95\textwidth]{example-image-a}}
    \end{minipage}
    \hfill
   \begin{minipage}{0.48\textwidth}
        \ifthenelse{\boolean{show_figs}}
        {\includegraphics[width=0.99\textwidth]{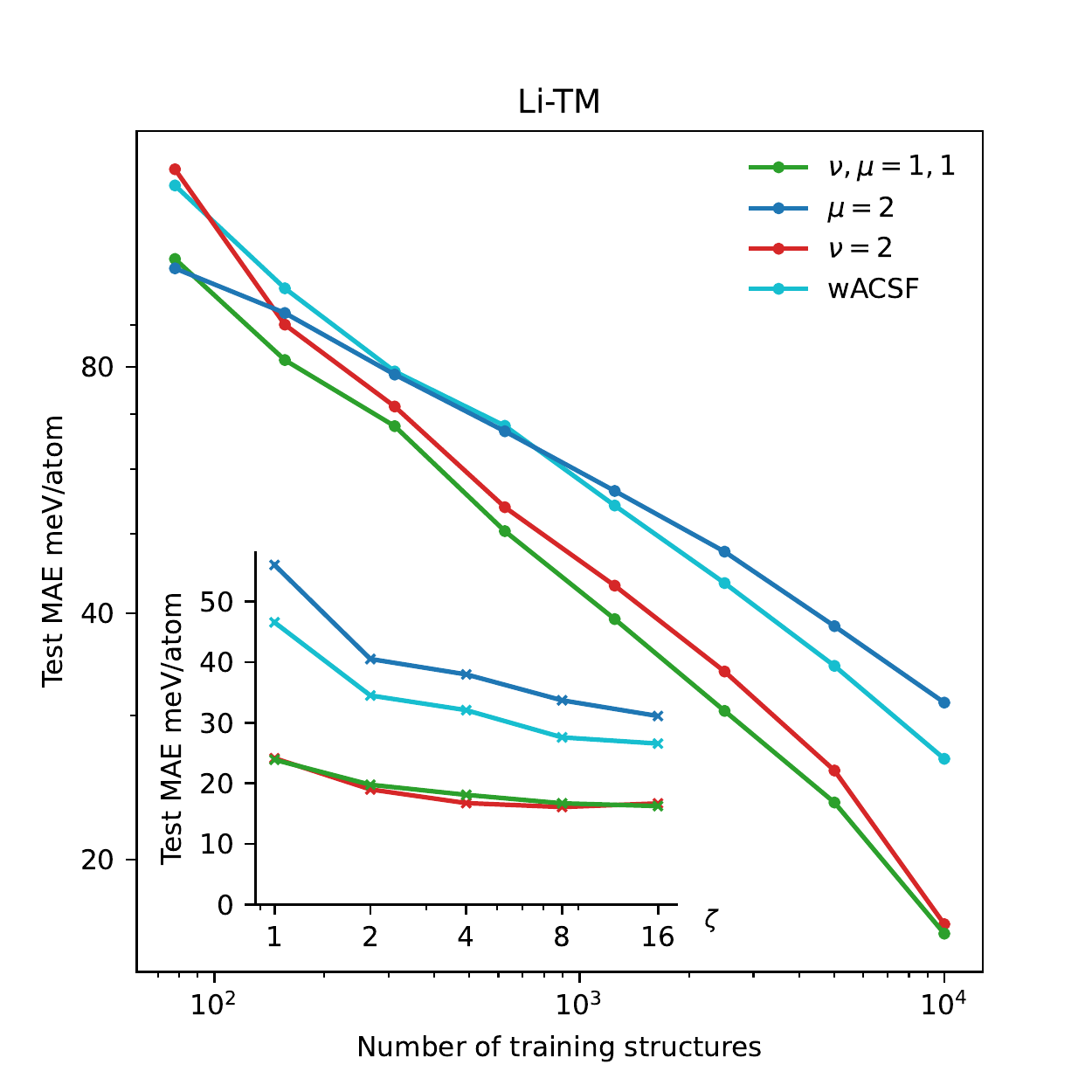}}
        {\includegraphics[width=0.95\textwidth]{example-image-a}}
    \end{minipage}
    \hfill
    \caption{Energy errors for KRR models with ACSF's as the atomic descriptors for the QM9 ($\zeta=2$) and Li-TM ($\zeta$=16 shown in main panel) datasets. The inset on the right shows how the error changed with $\zeta$ for a training set with $10^4$ configurations. }
  \label{fig:ACSF_fits}
\end{figure*}

Learning curves are shown in Figure \ref{fig:ACSF_fits}, where it can be seen that using the $\nu,\mu=1,1$ compression does not cause a noticeable decrease in model accuracy on either dataset. For the Li-TM dataset models were fit using $\zeta$=1, 2, 4, 8 and 16 to assess the effect of varying the flexibility of the model. As expected, the shorter descriptors, $\mu=2$ and wACSF, benefited more from increasing $\zeta$ but still never achieved comparable accuracy to the others. More interestingly, using $\nu,\mu=1,1$ provided the same accuracy as $\nu=2$ even with $\zeta=1$, despite the full descriptor being more than $5\times$ as long. These results clearly show how the ideas behind the generalized \ac{SOAP} kernel can be successfully applied to other body-ordered descriptors.

\section{Discussion}
As the number of elements $S$ increases, the length of many atomic descriptors increases drastically, with $S^\nu$ scaling common for $\nu+1$-body order descriptors. In this work we have sought non-data-driven way to reduce this scaling, with a focus on the SOAP power spectrum. We started by investigating the degree to which the density expansion coefficients can be recovered from the power spectrum. This analysis revealed that the power spectrum can be viewed as a collection of Gram matrices $P_l$, one for each total angular momentum number $l$, and that storing only a subset of components $W^TP_l$  is sufficient to preserve full information. This compression reduces the descriptor size from $\frac{1}{2}NS(NS+1)(L+1)$ to $NS(L+1)^2$, where $N$ and $L$ are the number of radial basis functions and highest order of spherical harmonic used in the density expansion respectively. Compressing the power spectrum in this way requires a single matrix of random weights $W$ to be stored and used consistently to compress all descriptors for a given dataset.

Next, we introduced the generalized SOAP kernel. The standard power spectrum, $\nu=2$, is a 3-body descriptor corresponding to a histogram of triangles, with one histogram for each pair of species.  In the generalised power spectrum the number of triangle vertices which are element-sensitive can be varied, as can the number of vertices which are projected onto the surface of the unit sphere. These modifications allow the scaling of the descriptor size with $S$ and $N$ to be set independently of the overall body-order and are applicable to all body-ordered descriptors. By again considering the ability to reconstruct the original power spectrum from the generalized one we showed that, subject to certain conditions, choosing $\nu,\mu^*=1,1$ (or $\nu,\mu=1,1$ where $N\ge2l+1$) does not lead to any loss of information. We also stress that descriptors based on the generalized kernel retain the element-wise sparsity of the original kernel, so that the number of non-zero components scales only with the number of elements present in a given environment, rather than total number present across a dataset.

The real-world performance of the compressions was tested using multiple numerical tests across a total of 5 pre-existing datasets. First, the information content, relative to the original power spectrum was analysed using the information imbalance approach of ref. \cite{glielmo2021ranking}. This analysis indicated that retaining element sensitivity on only one vertex was sufficient to ensure a minimal loss of information across all tested datasets. The constant complexity compression approaches performed well for the geometry optimised amino acids but incurred severe information loss on the liquid environments within the high-entropy alloy dataset. We believe this is because the atom type information for the QM9 and amino acid datasets is effectively encoded in the equilibrium geometry of the molecules, which suggests that such datasets are not well suited to assess the ability of a given descriptor to encode multi-element information. 

Secondly, models were fitted to the total energies for the QM9, Li-TM, amino acid and elpasolite datasets using linear and kernel ridge regression. The most promising compressions achieved very similar results to the full power spectrum across all datasets, whilst being significantly shorter. A notable deviation in behaviour was seen for the element-embedding fits to the elpasolite dataset, highlighting the differences between compression approaches. \ac{KRR} models were also fitted to the QM9 and Li-TM datasets using compressed versions of ACSF's, where, as before, the errors achieved with the $\nu,\mu=1,1$ compression was almost identical to those achieved with the full descriptor. Following this, \ac{KRR} models using the generalized \ac{SOAP} power spectrum were fitted to energies, forces and virials for the HEA dataset using the gap\_fit program \cite{gapfit}. The most accurate model that could be fitted, subject to practical restrictions on the quantity of training data and available memory, made use of the $\nu,\mu=1,1$ compression, providing concrete evidence that these compressions will be useful when fitting force-fields. Finally, the sensitivity of all descriptors to small perturbations was evaluated using the sensitivity matrix introduced in ref. \cite{parsaeifard2021assessment}. None of the compressions were found to significantly reduce the sensitivity of the descriptor, which is unsurprising given their relation to the single element power spectrum. 

We anticipate that these compressions will prove useful across a wide range of applications and that some of the ideas may be applicable to other body-ordered atomic descriptors, as was shown explicitly for \acp{ACSF}. In particular, the generalised kernel could provide  compression for approaches such as ACE \cite{drautz2019atomic, van2020regularised}, where the high body-orders, $\nu\geq4$, that are needed limit the number of different elements that can be treated. Furthermore, we stress that the compression ideas presented here can often be combined with pre-existing techniques, such as element-embedding, and that, in general, choosing the appropriate compression methods for a given situation is crucial. 

\section{Methods} \label{sec:methods}
\subsection{Datasets} \label{sec:datasets}
The datasets used for the numerical tests are:

\begin{enumerate}
    \item \textbf{QM9}: The QM9 dataset \cite{ramakrishnan2014quantum} contains $\sim$140,000 geometry optimized organic molecules containing only H, C, N, O and F. In this work we fit the internal energy at 0 K, reported as U0, and hereafter refer to this as energy.
    \item \textbf{HEA}: A quinary (W, Ta, Ni, Va and Mo) high entropy alloy dataset from ref. \cite{byggmastar2021modeling} containing 2329 configurations including bulk crystals, surfaces, vacancies, alloys and liquid structures. The training set was used for the distance-distance correlation, information imbalance and sensitivity analysis whilst the independent test from ref. \cite{byggmastar2021modeling} was used to assess the accuracy of the force-fields. 
    \item \textbf{Li-TM}: This dataset from ref. \cite{artrith2017efficient} consists of 16,407 Lithium Transition Metal Oxides formed from 11 differnet elements, (Li, Ti, O, Mn, Ni, Sc, V, Cr, Mn, Fe, Co and Cu).
    \item \textbf{Amino Acids}: A collection of 45,701 geometry optimised amino acids  from ref. \cite{ropo2016first} containing a total of 11 different elements. The dataset covers a total of 280 molecular systems - product of 20 proteingenic amino acids with 2 different backbone types (N-terminally acetylated or C-terminally amino-methylated) and 7 different cation additions (None, Ca$^{2+}$, Ba$^{2+}$, Sr$^{2+}$, Cd$^{2+}$, Pb$^{2+}$ or Hg$^{2+}$). 
    \item \textbf{Elpasolites}: A collection of $\sim$10,500 geometry optimised structures from ref. \cite{faber2016machine}. All main group elements up to Bi are present, 39 elements in total, and all structures share the elpasolite structural prototype. 
\end{enumerate}

\subsection{Element Embedding Parameters}
For the element embeddings, $d_J=2$ was used for most datasets with an additional test performed using $d_J=3$ for the elpasolite dataset. These embeddings were not optimised, as in refs. \cite{artrith2017efficient} and \cite{uhrin2021through}, and were both constructed using $w^1_{\alpha}=1$, so that the first density channel is element agnostic. For $d_J=2$ the weights for the second channel were chosen by ordering the elements according to atomic number and then assigning weights of 1,2,3,... so that for QM9 the $w^2_{\alpha}$ used were 1,2,3,4 and 5 for H,C,N,O and F respectively. For $d_J=3$ the weights for the second and third density channels were assigned using the group and period of each element in the periodic table, so that for sulphur $w_{1\alpha}=1$, $w_{2\alpha}=3$ and $w_{3\alpha}=16$. This was done in an attempt to capture the chemical similarity encoded in the periodic table and is similar to the encoding used in ref. \cite{faber2016machine}. For the elpasolite dataset the results of using optimised embeddings, denoted using $\bar{d_J}=2$ and $\bar{d_J}=4$, from ref. \cite{willatt2018feature} are also shown.

\subsection{Information Imbalance}\label{sec:methods_II}
The information imbalance is a way of measuring the relative information content of different distance measures. Whilst the distance-distance correlation compares all pairs of distances, the information imbalance is only concerned that the nearest neighbours of each environment are the same according to descriptor A and descriptor B.  More precisely, for each environment the distances to all other environments are computed using both distance A and distance B. These distances are then sorted and ranked from $0-N$ so that each environment has a rank according to A, $r_A$ and according to B, $r_B$. Then the information imbalance from A to B is defined as 
\begin{equation}
    \Delta_{A\rightarrow B} = \frac{2}{N} \langle r_A | r_B=1 \rangle
\end{equation}
where $N$ is the number of environments in the dataset and $\langle r_B | r_A=1 \rangle$ is the conditional average of $r_B$ given that $r_A=1$. Defined in this way $\Delta_{A \rightarrow B}$ is statistically confined to lie between 0, A contains the information in B, and 1, A is not informative about B. By comparing ranks, rather than distances, $\Delta_{A \rightarrow B}$ is insensitive to changes in scale and by considering only nearest neighbour distances $\Delta_{A \rightarrow B}$ is also well suited to handle non-linear relationships. Please refer to ref. \cite{glielmo2021ranking} for more details. 

\subsection{Sensitivity Matrix} \label{sec:methods_sensitivity}
Here we give a brief explanation of how the sensitivity matrix $\Lambda$ is constructed, please refer to ref. \cite{parsaeifard2021assessment} for full details. The distance $d$ between the original environment and the perturbed environment is given by 
\begin{equation*}
    d^2 = \sum_i \left( \Delta x_i \right)^2
\end{equation*}
where $\Delta x_i$ is the change in component $i$ of the descriptor $\boldsymbol{x}$. In terms of the atomic displacements this can be re-written as 
\begin{equation}
    d^2 = \sum_{jk} \Delta R_j \left( \sum_i \frac{\partial x_i}{\partial R_j} \frac{\partial x_i}{\partial R_k}  \right) \Delta R_k = \\
    \Delta\boldsymbol{ R}^T \mathbf{\Lambda}  \Delta\boldsymbol{ R}
\end{equation}

where $\Delta\boldsymbol{ R}$ is a vector of length $3N$ containing the small perturbations to the atomic positions. Defined as such, the distance between the original environment and one perturbed along an eigenvector $\boldsymbol{u}$ of $\Lambda$ is given by $d=\sqrt{\lambda}|\boldsymbol{u}|$ where $\Lambda \boldsymbol{u} =\lambda\boldsymbol{u}$. Thus, by examining the eigenvalues of $\Lambda$ we can detect if there are any perturbations that a given descriptor is insensitive to. We note that there will always be six zero eigenvalues, corresponding to three translations and three rotations, and that we expect additional zero eigenvalues for perturbations about symmetric atomic configurations \cite{pozdnyakov2021local}. This is demonstrated in Figure \ref{fig:sensitivity_overlay} where there are only 6 zero eigenvalues for the asymmetric liquid environments from the HEA dataset but many more zero eigenvalues for the symmetric environments found in the elpasolite dataset.

\subsection{KRR Models} \label{sec:models}
A simple linear model was used to fit the average chemical potential $\mu_\alpha$ for each element so that the predicted energy $\hat{E}_j$ for configuration $j$, denoted by $A_j$, was given by,
\begin{equation}
    \begin{split}
       \hat{E}_j &= \sum_{j\alpha} n_{j\alpha} \mu_\alpha + \boldsymbol{\beta}\cdot\boldsymbol{k}\left(A_j\right)
    \end{split}
    \label{eqn:energy_fit}
\end{equation}
where $n_{j\alpha}$ is the number of atoms of type $\alpha$ in $S_j$, $\boldsymbol{\beta}$ is a coefficient vector,  and $\boldsymbol{k}\left(A_j\right)$ is shorthand for the vector of kernels between $A_j$ and the structures in the training set,
$$\left[\boldsymbol{k}\left(A_j\right)\right]_i = k(A_i, A_j) =\sum_{\substack{x_j \in A_j \\ x_i \in A_i}} k(\mathbf{x}_i, \mathbf{x}_j)$$
where $\mathbf{x}_i$ is the descriptor of atom $i$, so that the kernel between structures is the sum over all pairwise atomic kernels. A polynomial kernel was used throughout so that $k(\boldsymbol{x}_i, \mathbf{x}_j) = \left(\mathbf{x}_i \cdot \mathbf{x}_j \right)^\zeta $ where $\zeta=1$ is equivalent to RR and $\zeta=2$, 4, 8 or 16 were used for KRR. We follow ref. \cite{deringer2021gaussian} in using the following loss function, motivated by the Gaussian Process Regression view,
$$ L = \left|\Sigma^{-1} (\mathbf{E} - \mathbf{\hat{E}})   \right|^2 + \boldsymbol{\beta}^TK\boldsymbol{\beta}    $$
where $K_{ij} = k(A_i, A_j)$, $\mathbf{E}$ is the vector of total energies for structures in the training set, $\mathbf{\hat{E}}$ are the predicted energies  and $\Sigma_{ij} = \sigma n_i \delta_{ij}$ where $n_i $ is the total number of atoms in $S_i$. Minimizing $L$ is equivalent to minimising the sum of the RMSE per atom on the training set and an $L_2$ regression penalty on $\boldsymbol{\beta}$. Doing so yields the following well known solution \cite{williams2006gaussian},
$$\boldsymbol{\beta} = \left(K + \Sigma\right)^{-1}\mathbf{E}  $$

In all cases multiple models were trained using a randomly selected train:test split, with all data not in the training set used as test data. The average MAE achieved across these models is reported with error bars indicating one standard deviation, whilst the regularisation strength $\sigma$ was chosen using k-fold cross validation, $k\sim10-20$.

\section{Data Availability}
All datasets used are freely available and the data used to generate all plots is available at \href{https://doi.org/10.5281/zenodo.5793851}{https://doi.org/10.5281/zenodo.5793851}.

\section{Code Availability}
The compressions outlined in Figure \ref{fig:3-body} have been implemented in the \ac{GAP} fitting code gap\_fit \cite{gapfit}. A Jupyter notebook demonstrating how the $W^TP_l$ compressed power spectrum is computed, and how the original power spectrum can be recovered from it, is available at \href{https://doi.org/10.5281/zenodo.5793851}{https://doi.org/10.5281/zenodo.5793851}.

\null\newpage
\bibliographystyle{naturemag}
\bibliography{arxiv.bib}

\begin{thebibliography}{10}
\expandafter\ifx\csname url\endcsname\relax
  \def\url#1{\texttt{#1}}\fi
\expandafter\ifx\csname urlprefix\endcsname\relax\def\urlprefix{URL }\fi
\providecommand{\bibinfo}[2]{#2}
\providecommand{\eprint}[2][]{\url{#2}}

\bibitem{burke2012perspective}
\bibinfo{author}{Burke, K.}
\newblock \bibinfo{title}{Perspective on density functional theory}.
\newblock \emph{\bibinfo{journal}{The Journal of chemical physics}}
  \textbf{\bibinfo{volume}{136}}, \bibinfo{pages}{150901}
  (\bibinfo{year}{2012}).

\bibitem{nightingale1998quantum}
\bibinfo{author}{Nightingale, M.~P.} \& \bibinfo{author}{Umrigar, C.~J.}
\newblock \emph{\bibinfo{title}{Quantum Monte Carlo methods in physics and
  chemistry}}.
\newblock \bibinfo{number}{525} (\bibinfo{publisher}{Springer Science \&
  Business Media}, \bibinfo{year}{1998}).

\bibitem{bartlett2007coupled}
\bibinfo{author}{Bartlett, R.~J.} \& \bibinfo{author}{Musia{\l}, M.}
\newblock \bibinfo{title}{Coupled-cluster theory in quantum chemistry}.
\newblock \emph{\bibinfo{journal}{Reviews of Modern Physics}}
  \textbf{\bibinfo{volume}{79}}, \bibinfo{pages}{291} (\bibinfo{year}{2007}).

\bibitem{behler2007generalized}
\bibinfo{author}{Behler, J.} \& \bibinfo{author}{Parrinello, M.}
\newblock \bibinfo{title}{Generalized neural-network representation of
  high-dimensional potential-energy surfaces}.
\newblock \emph{\bibinfo{journal}{Physical review letters}}
  \textbf{\bibinfo{volume}{98}}, \bibinfo{pages}{146401}
  (\bibinfo{year}{2007}).

\bibitem{bartok2010gaussian}
\bibinfo{author}{Bart{\'o}k, A.~P.}, \bibinfo{author}{Payne, M.~C.},
  \bibinfo{author}{Kondor, R.} \& \bibinfo{author}{Cs{\'a}nyi, G.}
\newblock \bibinfo{title}{Gaussian approximation potentials: The accuracy of
  quantum mechanics, without the electrons}.
\newblock \emph{\bibinfo{journal}{Physical review letters}}
  \textbf{\bibinfo{volume}{104}}, \bibinfo{pages}{136403}
  (\bibinfo{year}{2010}).

\bibitem{bartok2018machine}
\bibinfo{author}{Bart{\'o}k, A.~P.}, \bibinfo{author}{Kermode, J.},
  \bibinfo{author}{Bernstein, N.} \& \bibinfo{author}{Cs{\'a}nyi, G.}
\newblock \bibinfo{title}{Machine learning a general-purpose interatomic
  potential for silicon}.
\newblock \emph{\bibinfo{journal}{Physical Review X}}
  \textbf{\bibinfo{volume}{8}}, \bibinfo{pages}{041048} (\bibinfo{year}{2018}).

\bibitem{zuo2020performance}
\bibinfo{author}{Zuo, Y.} \emph{et~al.}
\newblock \bibinfo{title}{Performance and cost assessment of machine learning
  interatomic potentials}.
\newblock \emph{\bibinfo{journal}{The Journal of Physical Chemistry A}}
  \textbf{\bibinfo{volume}{124}}, \bibinfo{pages}{731--745}
  (\bibinfo{year}{2020}).

\bibitem{shapeev2016moment}
\bibinfo{author}{Shapeev, A.~V.}
\newblock \bibinfo{title}{Moment tensor potentials: A class of systematically
  improvable interatomic potentials}.
\newblock \emph{\bibinfo{journal}{Multiscale Modeling \& Simulation}}
  \textbf{\bibinfo{volume}{14}}, \bibinfo{pages}{1153--1173}
  (\bibinfo{year}{2016}).

\bibitem{van2020regularised}
\bibinfo{author}{van~der Oord, C.}, \bibinfo{author}{Dusson, G.},
  \bibinfo{author}{Cs{\'a}nyi, G.} \& \bibinfo{author}{Ortner, C.}
\newblock \bibinfo{title}{Regularised atomic body-ordered permutation-invariant
  polynomials for the construction of interatomic potentials}.
\newblock \emph{\bibinfo{journal}{Machine Learning: Science and Technology}}
  \textbf{\bibinfo{volume}{1}}, \bibinfo{pages}{015004} (\bibinfo{year}{2020}).

\bibitem{faber2018alchemical}
\bibinfo{author}{Faber, F.~A.}, \bibinfo{author}{Christensen, A.~S.},
  \bibinfo{author}{Huang, B.} \& \bibinfo{author}{Von~Lilienfeld, O.~A.}
\newblock \bibinfo{title}{Alchemical and structural distribution based
  representation for universal quantum machine learning}.
\newblock \emph{\bibinfo{journal}{The Journal of chemical physics}}
  \textbf{\bibinfo{volume}{148}}, \bibinfo{pages}{241717}
  (\bibinfo{year}{2018}).

\bibitem{deringer2021origins}
\bibinfo{author}{Deringer, V.~L.} \emph{et~al.}
\newblock \bibinfo{title}{Origins of structural and electronic transitions in
  disordered silicon}.
\newblock \emph{\bibinfo{journal}{Nature}} \textbf{\bibinfo{volume}{589}},
  \bibinfo{pages}{59--64} (\bibinfo{year}{2021}).

\bibitem{jain2016research}
\bibinfo{author}{Jain, A.}, \bibinfo{author}{Persson, K.~A.} \&
  \bibinfo{author}{Ceder, G.}
\newblock \bibinfo{title}{Research update: The materials genome initiative:
  Data sharing and the impact of collaborative ab initio databases}.
\newblock \emph{\bibinfo{journal}{APL Materials}} \textbf{\bibinfo{volume}{4}},
  \bibinfo{pages}{053102} (\bibinfo{year}{2016}).

\bibitem{saal2013materials}
\bibinfo{author}{Saal, J.~E.}, \bibinfo{author}{Kirklin, S.},
  \bibinfo{author}{Aykol, M.}, \bibinfo{author}{Meredig, B.} \&
  \bibinfo{author}{Wolverton, C.}
\newblock \bibinfo{title}{Materials design and discovery with high-throughput
  density functional theory: the open quantum materials database (oqmd)}.
\newblock \emph{\bibinfo{journal}{Jom}} \textbf{\bibinfo{volume}{65}},
  \bibinfo{pages}{1501--1509} (\bibinfo{year}{2013}).

\bibitem{draxl2018nomad}
\bibinfo{author}{Draxl, C.} \& \bibinfo{author}{Scheffler, M.}
\newblock \bibinfo{title}{Nomad: The fair concept for big data-driven materials
  science}.
\newblock \emph{\bibinfo{journal}{Mrs Bulletin}} \textbf{\bibinfo{volume}{43}},
  \bibinfo{pages}{676--682} (\bibinfo{year}{2018}).

\bibitem{bernstein2019novo}
\bibinfo{author}{Bernstein, N.}, \bibinfo{author}{Cs{\'a}nyi, G.} \&
  \bibinfo{author}{Deringer, V.~L.}
\newblock \bibinfo{title}{De novo exploration and self-guided learning of
  potential-energy surfaces}.
\newblock \emph{\bibinfo{journal}{npj Computational Materials}}
  \textbf{\bibinfo{volume}{5}}, \bibinfo{pages}{1--9} (\bibinfo{year}{2019}).

\bibitem{musil2021physics}
\bibinfo{author}{Musil, F.} \emph{et~al.}
\newblock \bibinfo{title}{Physics-inspired structural representations for
  molecules and materials}.
\newblock \emph{\bibinfo{journal}{Chemical Reviews}}
  \textbf{\bibinfo{volume}{121}}, \bibinfo{pages}{9759--9815}
  (\bibinfo{year}{2021}).

\bibitem{bartok2013representing}
\bibinfo{author}{Bart{\'o}k, A.~P.}, \bibinfo{author}{Kondor, R.} \&
  \bibinfo{author}{Cs{\'a}nyi, G.}
\newblock \bibinfo{title}{On representing chemical environments}.
\newblock \emph{\bibinfo{journal}{Physical Review B}}
  \textbf{\bibinfo{volume}{87}}, \bibinfo{pages}{184115}
  (\bibinfo{year}{2013}).

\bibitem{artrith2017efficient}
\bibinfo{author}{Artrith, N.}, \bibinfo{author}{Urban, A.} \&
  \bibinfo{author}{Ceder, G.}
\newblock \bibinfo{title}{Efficient and accurate machine-learning interpolation
  of atomic energies in compositions with many species}.
\newblock \emph{\bibinfo{journal}{Physical Review B}}
  \textbf{\bibinfo{volume}{96}}, \bibinfo{pages}{014112}
  (\bibinfo{year}{2017}).

\bibitem{uhrin2021through}
\bibinfo{author}{Uhrin, M.}
\newblock \bibinfo{title}{Through the eyes of a descriptor: Constructing
  complete, invertible, descriptions of atomic environments}.
\newblock \emph{\bibinfo{journal}{arXiv preprint arXiv:2104.09319}}
  (\bibinfo{year}{2021}).

\bibitem{schutt2018schnet}
\bibinfo{author}{Sch{\"u}tt, K.~T.}, \bibinfo{author}{Sauceda, H.~E.},
  \bibinfo{author}{Kindermans, P.-J.}, \bibinfo{author}{Tkatchenko, A.} \&
  \bibinfo{author}{M{\"u}ller, K.-R.}
\newblock \bibinfo{title}{Schnet--a deep learning architecture for molecules
  and materials}.
\newblock \emph{\bibinfo{journal}{The Journal of Chemical Physics}}
  \textbf{\bibinfo{volume}{148}}, \bibinfo{pages}{241722}
  (\bibinfo{year}{2018}).

\bibitem{willatt2018feature}
\bibinfo{author}{Willatt, M.~J.}, \bibinfo{author}{Musil, F.} \&
  \bibinfo{author}{Ceriotti, M.}
\newblock \bibinfo{title}{Feature optimization for atomistic machine learning
  yields a data-driven construction of the periodic table of the elements}.
\newblock \emph{\bibinfo{journal}{Physical Chemistry Chemical Physics}}
  \textbf{\bibinfo{volume}{20}}, \bibinfo{pages}{29661--29668}
  (\bibinfo{year}{2018}).

\bibitem{glielmo2021ranking}
\bibinfo{author}{Glielmo, A.}, \bibinfo{author}{Zeni, C.},
  \bibinfo{author}{Cheng, B.}, \bibinfo{author}{Csanyi, G.} \&
  \bibinfo{author}{Laio, A.}
\newblock \bibinfo{title}{Ranking the information content of distance
  measures}.
\newblock \emph{\bibinfo{journal}{arXiv preprint arXiv:2104.15079}}
  (\bibinfo{year}{2021}).

\bibitem{onat2020sensitivity}
\bibinfo{author}{Onat, B.}, \bibinfo{author}{Ortner, C.} \&
  \bibinfo{author}{Kermode, J.~R.}
\newblock \bibinfo{title}{Sensitivity and dimensionality of atomic environment
  representations used for machine learning interatomic potentials}.
\newblock \emph{\bibinfo{journal}{The Journal of Chemical Physics}}
  \textbf{\bibinfo{volume}{153}}, \bibinfo{pages}{144106}
  (\bibinfo{year}{2020}).

\bibitem{mahoney2009cur}
\bibinfo{author}{Mahoney, M.~W.} \& \bibinfo{author}{Drineas, P.}
\newblock \bibinfo{title}{Cur matrix decompositions for improved data
  analysis}.
\newblock \emph{\bibinfo{journal}{Proceedings of the National Academy of
  Sciences}} \textbf{\bibinfo{volume}{106}}, \bibinfo{pages}{697--702}
  (\bibinfo{year}{2009}).

\bibitem{nigam2020recursive}
\bibinfo{author}{Nigam, J.}, \bibinfo{author}{Pozdnyakov, S.} \&
  \bibinfo{author}{Ceriotti, M.}
\newblock \bibinfo{title}{Recursive evaluation and iterative contraction of
  n-body equivariant features}.
\newblock \emph{\bibinfo{journal}{The Journal of Chemical Physics}}
  \textbf{\bibinfo{volume}{153}}, \bibinfo{pages}{121101}
  (\bibinfo{year}{2020}).

\bibitem{goscinski2021optimal}
\bibinfo{author}{Goscinski, A.}, \bibinfo{author}{Musil, F.},
  \bibinfo{author}{Pozdnyakov, S.}, \bibinfo{author}{Nigam, J.} \&
  \bibinfo{author}{Ceriotti, M.}
\newblock \bibinfo{title}{Optimal radial basis for density-based atomic
  representations}.
\newblock \emph{\bibinfo{journal}{The Journal of Chemical Physics}}
  \textbf{\bibinfo{volume}{155}}, \bibinfo{pages}{104106}
  (\bibinfo{year}{2021}).

\bibitem{draxl2019nomad}
\bibinfo{author}{Draxl, C.} \& \bibinfo{author}{Scheffler, M.}
\newblock \bibinfo{title}{The nomad laboratory: from data sharing to artificial
  intelligence}.
\newblock \emph{\bibinfo{journal}{Journal of Physics: Materials}}
  \textbf{\bibinfo{volume}{2}}, \bibinfo{pages}{036001} (\bibinfo{year}{2019}).

\bibitem{quippy}
\bibinfo{author}{Cs{\'a}nyi, G.} \emph{et~al.}
\newblock \bibinfo{title}{Expressive programming for computational physics in
  fortran 95+}.
\newblock \emph{\bibinfo{journal}{IoP Comput. Phys. Newsletter}}
  \bibinfo{pages}{Spring 2007} (\bibinfo{year}{2007}).

\bibitem{Kermode2020-wu}
\bibinfo{author}{Kermode, J.~R.}
\newblock \bibinfo{title}{f90wrap: an automated tool for constructing deep
  python interfaces to modern fortran codes}.
\newblock \emph{\bibinfo{journal}{J. Phys. Condens. Matter}}
  (\bibinfo{year}{2020}).

\bibitem{dscribe}
\bibinfo{author}{Himanen, L.} \emph{et~al.}
\newblock \bibinfo{title}{{DScribe: Library of descriptors for machine learning
  in materials science}}.
\newblock \emph{\bibinfo{journal}{Computer Physics Communications}}
  \textbf{\bibinfo{volume}{247}}, \bibinfo{pages}{106949}
  (\bibinfo{year}{2020}).
\newblock \urlprefix\url{https://doi.org/10.1016/j.cpc.2019.106949}.

\bibitem{musil2021efficient}
\bibinfo{author}{Musil, F.} \emph{et~al.}
\newblock \bibinfo{title}{Efficient implementation of atom-density
  representations}.
\newblock \emph{\bibinfo{journal}{The Journal of Chemical Physics}}
  \textbf{\bibinfo{volume}{154}}, \bibinfo{pages}{114109}
  (\bibinfo{year}{2021}).

\bibitem{behler2011atom}
\bibinfo{author}{Behler, J.}
\newblock \bibinfo{title}{Atom-centered symmetry functions for constructing
  high-dimensional neural network potentials}.
\newblock \emph{\bibinfo{journal}{The Journal of chemical physics}}
  \textbf{\bibinfo{volume}{134}}, \bibinfo{pages}{074106}
  (\bibinfo{year}{2011}).

\bibitem{de2016comparing}
\bibinfo{author}{De, S.}, \bibinfo{author}{Bart{\'o}k, A.~P.},
  \bibinfo{author}{Cs{\'a}nyi, G.} \& \bibinfo{author}{Ceriotti, M.}
\newblock \bibinfo{title}{Comparing molecules and solids across structural and
  alchemical space}.
\newblock \emph{\bibinfo{journal}{Physical Chemistry Chemical Physics}}
  \textbf{\bibinfo{volume}{18}}, \bibinfo{pages}{13754--13769}
  (\bibinfo{year}{2016}).

\bibitem{riley1999mathematical}
\bibinfo{author}{Riley, K.~F.}, \bibinfo{author}{Hobson, M.~P.} \&
  \bibinfo{author}{Bence, S.~J.}
\newblock \bibinfo{title}{Mathematical methods for physics and engineering}
  (\bibinfo{year}{1999}).

\bibitem{caro2019optimizing}
\bibinfo{author}{Caro, M.~A.}
\newblock \bibinfo{title}{Optimizing many-body atomic descriptors for enhanced
  computational performance of machine learning based interatomic potentials}.
\newblock \emph{\bibinfo{journal}{Physical Review B}}
  \textbf{\bibinfo{volume}{100}}, \bibinfo{pages}{024112}
  (\bibinfo{year}{2019}).

\bibitem{turbogap_compress}
\bibinfo{author}{Caro, M.~A.}
\newblock \bibinfo{title}{Turbogap compression}.
\newblock
  \urlprefix\url{https://github.com/mcaroba/turbogap/tree/master/tools/compress_indices}.

\bibitem{pozdnyakov2020incompleteness}
\bibinfo{author}{Pozdnyakov, S.~N.} \emph{et~al.}
\newblock \bibinfo{title}{Incompleteness of atomic structure representations}.
\newblock \emph{\bibinfo{journal}{Physical Review Letters}}
  \textbf{\bibinfo{volume}{125}}, \bibinfo{pages}{166001}
  (\bibinfo{year}{2020}).

\bibitem{deringer2020general}
\bibinfo{author}{Deringer, V.~L.}, \bibinfo{author}{Caro, M.~A.} \&
  \bibinfo{author}{Cs{\'a}nyi, G.}
\newblock \bibinfo{title}{A general-purpose machine-learning force field for
  bulk and nanostructured phosphorus}.
\newblock \emph{\bibinfo{journal}{Nature communications}}
  \textbf{\bibinfo{volume}{11}}, \bibinfo{pages}{1--11} (\bibinfo{year}{2020}).

\bibitem{bartok2017machine}
\bibinfo{author}{Bart{\'o}k, A.~P.} \emph{et~al.}
\newblock \bibinfo{title}{Machine learning unifies the modeling of materials
  and molecules}.
\newblock \emph{\bibinfo{journal}{Science advances}}
  \textbf{\bibinfo{volume}{3}}, \bibinfo{pages}{e1701816}
  (\bibinfo{year}{2017}).

\bibitem{artrith2016implementation}
\bibinfo{author}{Artrith, N.} \& \bibinfo{author}{Urban, A.}
\newblock \bibinfo{title}{An implementation of artificial neural-network
  potentials for atomistic materials simulations: Performance for tio2}.
\newblock \emph{\bibinfo{journal}{Computational Materials Science}}
  \textbf{\bibinfo{volume}{114}}, \bibinfo{pages}{135--150}
  (\bibinfo{year}{2016}).

\bibitem{drautz2019atomic}
\bibinfo{author}{Drautz, R.}
\newblock \bibinfo{title}{Atomic cluster expansion for accurate and
  transferable interatomic potentials}.
\newblock \emph{\bibinfo{journal}{Physical Review B}}
  \textbf{\bibinfo{volume}{99}}, \bibinfo{pages}{014104}
  (\bibinfo{year}{2019}).

\bibitem{gastegger2018wacsf}
\bibinfo{author}{Gastegger, M.}, \bibinfo{author}{Schwiedrzik, L.},
  \bibinfo{author}{Bittermann, M.}, \bibinfo{author}{Berzsenyi, F.} \&
  \bibinfo{author}{Marquetand, P.}
\newblock \bibinfo{title}{wacsf—weighted atom-centered symmetry functions as
  descriptors in machine learning potentials}.
\newblock \emph{\bibinfo{journal}{The Journal of chemical physics}}
  \textbf{\bibinfo{volume}{148}}, \bibinfo{pages}{241709}
  (\bibinfo{year}{2018}).

\bibitem{batzner2021se}
\bibinfo{author}{Batzner, S.} \emph{et~al.}
\newblock \bibinfo{title}{Se (3)-equivariant graph neural networks for
  data-efficient and accurate interatomic potentials}.
\newblock \emph{\bibinfo{journal}{arXiv preprint arXiv:2101.03164}}
  (\bibinfo{year}{2021}).

\bibitem{anderson2019cormorant}
\bibinfo{author}{Anderson, B.}, \bibinfo{author}{Hy, T.-S.} \&
  \bibinfo{author}{Kondor, R.}
\newblock \bibinfo{title}{Cormorant: Covariant molecular neural networks}.
\newblock \emph{\bibinfo{journal}{arXiv preprint arXiv:1906.04015}}
  (\bibinfo{year}{2019}).

\bibitem{e3nn}
\bibinfo{author}{Geiger, M.} \emph{et~al.}
\newblock \bibinfo{title}{Euclidean neural networks: e3nn}
  (\bibinfo{year}{2020}).
\newblock \urlprefix\url{https://doi.org/10.5281/zenodo.5292912}.

\bibitem{weiler20183d}
\bibinfo{author}{Weiler, M.}, \bibinfo{author}{Geiger, M.},
  \bibinfo{author}{Welling, M.}, \bibinfo{author}{Boomsma, W.} \&
  \bibinfo{author}{Cohen, T.}
\newblock \bibinfo{title}{3d steerable cnns: Learning rotationally equivariant
  features in volumetric data}.
\newblock \emph{\bibinfo{journal}{arXiv preprint arXiv:1807.02547}}
  (\bibinfo{year}{2018}).

\bibitem{thomas2018tensor}
\bibinfo{author}{Thomas, N.} \emph{et~al.}
\newblock \bibinfo{title}{Tensor field networks: Rotation-and
  translation-equivariant neural networks for 3d point clouds}.
\newblock \emph{\bibinfo{journal}{arXiv preprint arXiv:1802.08219}}
  (\bibinfo{year}{2018}).

\bibitem{parsaeifard2021assessment}
\bibinfo{author}{Parsaeifard, B.} \emph{et~al.}
\newblock \bibinfo{title}{An assessment of the structural resolution of various
  fingerprints commonly used in machine learning}.
\newblock \emph{\bibinfo{journal}{Machine Learning: Science and Technology}}
  \textbf{\bibinfo{volume}{2}}, \bibinfo{pages}{015018} (\bibinfo{year}{2021}).

\bibitem{calsaverini2009information}
\bibinfo{author}{Calsaverini, R.~S.} \& \bibinfo{author}{Vicente, R.}
\newblock \bibinfo{title}{An information-theoretic approach to statistical
  dependence: Copula information}.
\newblock \emph{\bibinfo{journal}{EPL (Europhysics Letters)}}
  \textbf{\bibinfo{volume}{88}}, \bibinfo{pages}{68003} (\bibinfo{year}{2009}).

\bibitem{gapfit}
\bibinfo{title}{The gap\_fit code}.
\newblock \urlprefix\url{https://github.com/libAtoms/GAP}.

\bibitem{byggmastar2021modeling}
\bibinfo{author}{Byggmastar, J.}, \bibinfo{author}{Nordlund, K.} \&
  \bibinfo{author}{Djurabekova, F.}
\newblock \bibinfo{title}{Modeling refractory high-entropy alloys with
  efficient machine-learned interatomic potentials: defects and segregation}
  (\bibinfo{year}{2021}).
\newblock \eprint{2106.03369}.

\bibitem{pozdnyakov2021local}
\bibinfo{author}{Pozdnyakov, S.~N.}, \bibinfo{author}{Zhang, L.},
  \bibinfo{author}{Ortner, C.}, \bibinfo{author}{Cs{\'a}nyi, G.} \&
  \bibinfo{author}{Ceriotti, M.}
\newblock \bibinfo{title}{Local invertibility and sensitivity of atomic
  structure-feature mappings}.
\newblock \emph{\bibinfo{journal}{arXiv preprint arXiv:2109.11440}}
  (\bibinfo{year}{2021}).

\bibitem{ramakrishnan2014quantum}
\bibinfo{author}{Ramakrishnan, R.}, \bibinfo{author}{Dral, P.~O.},
  \bibinfo{author}{Rupp, M.} \& \bibinfo{author}{von Lilienfeld, O.~A.}
\newblock \bibinfo{title}{Quantum chemistry structures and properties of 134
  kilo molecules}.
\newblock \emph{\bibinfo{journal}{Scientific Data}}
  \textbf{\bibinfo{volume}{1}} (\bibinfo{year}{2014}).

\bibitem{ropo2016first}
\bibinfo{author}{Ropo, M.}, \bibinfo{author}{Schneider, M.},
  \bibinfo{author}{Baldauf, C.} \& \bibinfo{author}{Blum, V.}
\newblock \bibinfo{title}{First-principles data set of 45,892 isolated and
  cation-coordinated conformers of 20 proteinogenic amino acids}.
\newblock \emph{\bibinfo{journal}{Scientific data}}
  \textbf{\bibinfo{volume}{3}}, \bibinfo{pages}{1--13} (\bibinfo{year}{2016}).

\bibitem{faber2016machine}
\bibinfo{author}{Faber, F.~A.}, \bibinfo{author}{Lindmaa, A.},
  \bibinfo{author}{Von~Lilienfeld, O.~A.} \& \bibinfo{author}{Armiento, R.}
\newblock \bibinfo{title}{Machine learning energies of 2 million elpasolite (a
  b c 2 d 6) crystals}.
\newblock \emph{\bibinfo{journal}{Physical review letters}}
  \textbf{\bibinfo{volume}{117}}, \bibinfo{pages}{135502}
  (\bibinfo{year}{2016}).

\bibitem{deringer2021gaussian}
\bibinfo{author}{Deringer, V.~L.} \emph{et~al.}
\newblock \bibinfo{title}{Gaussian process regression for materials and
  molecules}.
\newblock \emph{\bibinfo{journal}{Chemical Reviews}}  (\bibinfo{year}{2021}).

\bibitem{williams2006gaussian}
\bibinfo{author}{Williams, C.~K.} \& \bibinfo{author}{Rasmussen, C.~E.}
\newblock \emph{\bibinfo{title}{Gaussian processes for machine learning}},
  vol.~\bibinfo{volume}{2} (\bibinfo{publisher}{MIT press Cambridge, MA},
  \bibinfo{year}{2006}).

\bibitem{kaufmann1989single}
\bibinfo{author}{Kaufmann, K.} \& \bibinfo{author}{Baumeister, W.}
\newblock \bibinfo{title}{Single-centre expansion of gaussian basis functions
  and the angular decomposition of their overlap integrals}.
\newblock \emph{\bibinfo{journal}{Journal of Physics B: Atomic, Molecular and
  Optical Physics}} \textbf{\bibinfo{volume}{22}}, \bibinfo{pages}{1}
  (\bibinfo{year}{1989}).

\end{thebibliography}

\section{Acknowledgements}
The authors would like to thank Luca Ghiringhelli, Ioan-Bogdan Magdău and Albert Bartók-Pártay for helpful discussions. We acknowledge support from the NOMAD Centre of Excellence, funded by the European Commission under grant agreement 951786. JRK acknowledges additional support provided by the Leverhulme Trust under grant RPG-2017-191. We are grateful for computational support from the UK national high performance computing service, ARCHER, for which access was obtained via the UKCP consortium and funded by EPSRC grant reference EP/P022065/1. 

\section{Author Contributions}
JRK and GC jointly designed the research. JPD developed the compression strategies, performed the calculations and wrote the paper, with input from JRK and GC at all stages. All authors revised the paper and approved its final version.

\section{Competing Interests}
The authors declare no competing interests.

%\include{SI}

%%%%%%%%%% Merge with supplemental materials %%%%%%%%%%
\clearpage
\widetext
\begin{center}
\textbf{\large \hypertarget{SI}{Supporting Information} }
\end{center}
\setcounter{figure}{0}
\renewcommand{\theequation}{S\arabic{equation}}
\renewcommand{\thefigure}{S\arabic{figure}}
\renewcommand{\theHfigure}{Supplement.\thefigure}

\subsection{Parallel density expansion vectors for single neighbour}

The density corresponding to a single neighbour atom $\rho^i(\mathbf{r})$ at position $\mathbf{r}_i$ can be expanded as \cite{kaufmann1989single} 

\begin{equation*}
\begin{split}
    \rho^i(\boldsymbol{r}) &= \exp{\left[-\alpha \left| \mathbf{r}-\mathbf{r}_i\right|^2\right]} \\
    &= 4\pi \exp{\left[-\alpha(r-r_i) ^2\right]} 
    \sum_{lm} l_l(2\alpha rr_i) Y_{lm}(\mathbf{\hat{r}}) Y^*_{lm}(\mathbf{\hat{r}}_i)
\end{split}
\end{equation*}

where $\alpha = \frac{1}{2\sigma^2}$ sets the width of the Gaussian used in the density expansion and $l_l$ is a modified spherical Bessel function of the first kind. Using the orthogonality relations of the radial basis functions $g_n(r)$ and the spherical harmonics $Y_{lm}(\mathbf{\hat{r}})$ the density expansion coefficients corresponding to the single neighbour atom $c^i_{nlm}$ can be written as 

\begin{equation*}
    c^i_{nlm} = \int  r^2 dr d\mathbf{\hat{r}} \rho^i(\mathbf{r}) g_n(r) Y^*_{lm}(\mathbf{\hat{r}})
\end{equation*}

which evaluates to 

\begin{equation*}
\begin{split}
    c^i_{nlm} = 4\pi \sum_{l'm'} Y^*_{l'm'}(\mathbf{\hat{r}}_i) 
    \underbrace{\int r^2 dr  \exp{\left[-\alpha(r-r_i) ^2\right]} 
     l_{l'}(2\alpha rr_i) g_n(r)}_{b_{nl'}(r_i)} \\
     \underbrace{\int d\mathbf{\hat{r}} Y^*_{lm}(\mathbf{\hat{r}}) Y_{l'm'}(\mathbf{\hat{r}})}_{\delta_{ll'}\delta_{mm'}}
\end{split}
\end{equation*}

\begin{equation*}
    c^i_{nlm} = 4\pi\  b_{nl}(r_i) Y^*_{lm}(\mathbf{\hat{r}}_i)
\end{equation*}

where $b_{nl}(r_i)$ depends on $n$, $l$ and the distance to the neighbour atom $r_i$. Consequently the ratio between $c^i_{nlm}$ and $c^i_{nlm'}$ is independent of $n$, so that $\mathbf{c}^i_{nl}$ and $\mathbf{c}^i_{n'l}$ are parallel.

\begin{equation*}
    \frac{c^i_{nlm}}{c^i_{nlm'}} = \frac{ Y^*_{lm}(\mathbf{\hat{r}}_i)}{Y^*_{lm'}(\mathbf{\hat{r}}_i)}
\end{equation*}

\subsection{Real vs Complex Spherical Harmonics}

$$X_l = (\mathbf{c}^\alpha_1, \mathbf{c}^\alpha_2, \dots,\mathbf{c}^\beta_1,\mathbf{c}^\beta_2, \dots,  \mathbf{c}^S_N)$$

$$\text{dim}(\mathbf{c}^\alpha_n) = 2l+1$$

\noindent That the rank$(X_l)\leq 2l+1$ is clear \textbf{if} the real spherical harmonics are used in the density expansion, so that the $c^\alpha_{nlm}$ are also real. The same also holds if the complex spherical harmonics are used because $Y_{lm}^*(\mathbf{\hat{r}}) = (-1)^mY_{l-m}(\mathbf{\hat{r}})$, which implies that $c_{nl0} = c^*_{nl0}$ and $c_{nl-m} = (-1)^m c^*_{lm}$. This means that the $\mathbf{c}_{nl}$ still span a $2l+1$ dimensional space, rather than the full $2\cdot(2l+1)$ dimensional space that would be spanned if the density expansion coefficients were independent (as would occur if the density was complex). 

\clearpage
\subsection{Supplementary Figures}

\begin{figure}[h!]
  \includegraphics[width=0.45\textwidth]{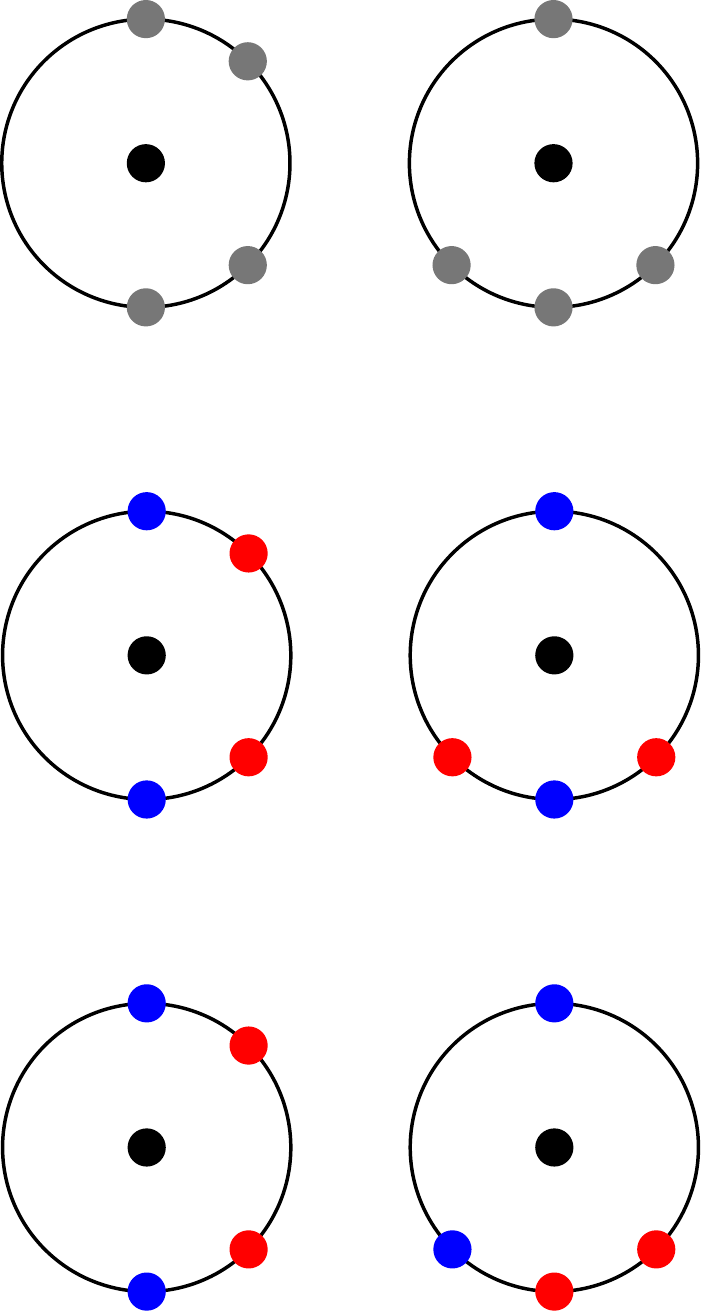}
  \caption{Three pairs of environments are show. The top row shows a pair of distinct environments that are not distinguished by the SOAP power spectrum - taken from ref. \cite{pozdnyakov2020incompleteness}. The middle and bottom rows show two possible variations using two differnet elements. The ``coloring'' on the bottom row breaks the degeneracy for $\nu=2$ (and $\nu,\mu=1,1$) whilst that shown in the middle row does not. There is no coloring of this pair that breaks the degeneracy for $\nu=2$ but not $\nu,\mu=1,1$. }
  \label{fig:bonus_degen}
\end{figure}

\begin{figure}[h!]
    \begin{minipage}{0.5\textwidth}
      \ifthenelse{\boolean{show_figs}}
        {\includegraphics[width=0.99\textwidth]{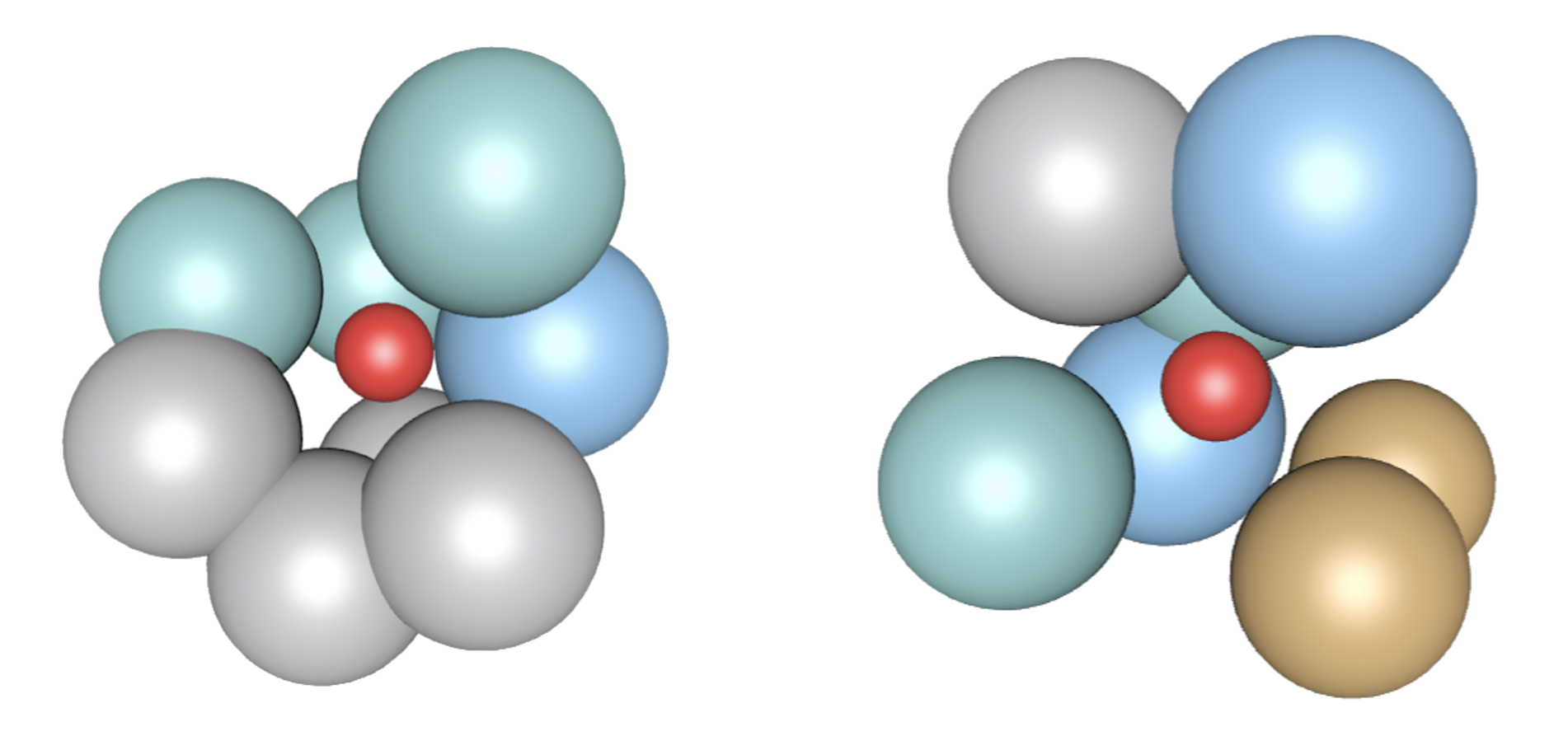}}
        {\includegraphics[width=0.95\textwidth]{example-image-a}}

    \end{minipage}
    \hfill
    \begin{minipage}{0.3\textwidth}
	\begin{tabular}{ccc}
	   \hline
	    Type & Length & d \\%& $k(\rho, \rho')$	\\
		\hline
		$\nu=2$ & $11760$ & 1.07 \\%& 0.430\\
		$\nu,\mu=1,1$ & 5760  & 0.856 \\%& 0.634 \\
		$\nu,\mu^*=1,1$ & 4800  & 0.849 \\%& 0.639 \\
		$\nu,\hat{\mu}=1,1$  & 480 & 0.856 \\%& 0.633 \\
		$\nu,d_J=2,2$ & 3000  & 0.106 \\%& 0.994 \\
		$W^TP_l$ & 4800 & $0.91\pm0.02$  \\ 
		\hline
	\end{tabular}	
    \end{minipage}\hfill
  \caption{Two environments, modified from the HEA dataset, containing V (grey), Ta (blue), Mo (green) and Nb (gold) are shown. The central atom is marked in red for clarity and the central weight has been set to 0 so that only the neighbour atoms affect the distance between the environments. The density expansion was performed using $N=12$, $L=9$, $r_{cut}=3.2$ \AA, $\sigma=0.4$ \AA\space and a cut-off transition width of 0.1 \AA. All of the neighbouring atoms are within $3$ \AA\space of the central atom so there are no cut-off effects present. The maximum distance across the dataset for all descriptors was $\approx\sqrt{2}$ so that a direct comparison between distances is meaningful. The distances between the two environments are shown in the table on the right; for $W^TP_l$ the mean and standard deviation from 20 sets of randomly chosen weights are shown. }
  \label{fig:HEA_degen}
\end{figure}

\begin{figure}[h!]
  \includegraphics[width=0.6\textwidth]{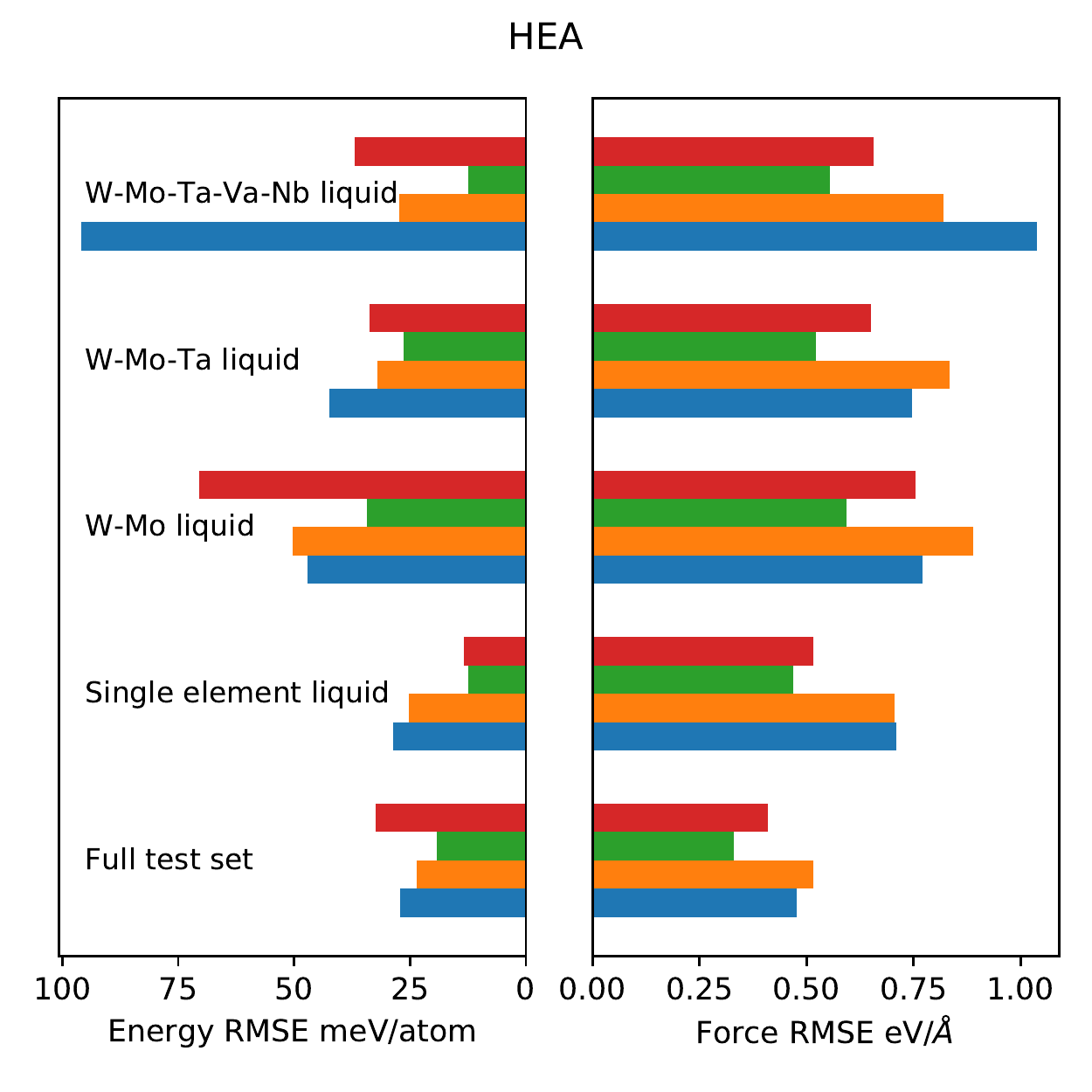}
  \caption{The energy and force errors on the test set, stratified by configuration type, are shown for the 2b+3b (blue) and 2b+SOAP models, $\nu,\hat{\mu}=1,1$ (orange), $\nu,\mu=1,1$ (green) and $\nu=2$ (red). Note the big improvement in the energy errors for the quinary alloy liquid. That the $\nu,\mu=1,1$ model achieves lower errors on the quinary liquid than the simpler W-Mo and W-Mo-Ta liquids can be attributed to a lack of training data; there are 55 training configurations for the former whilst for the others there are only 3. }
  \label{fig:HEA_errors}
\end{figure}

\begin{figure*}[h!]

    \begin{minipage}{0.48\textwidth}
        \ifthenelse{\boolean{show_figs}}
        {\includegraphics[width=0.99\textwidth]{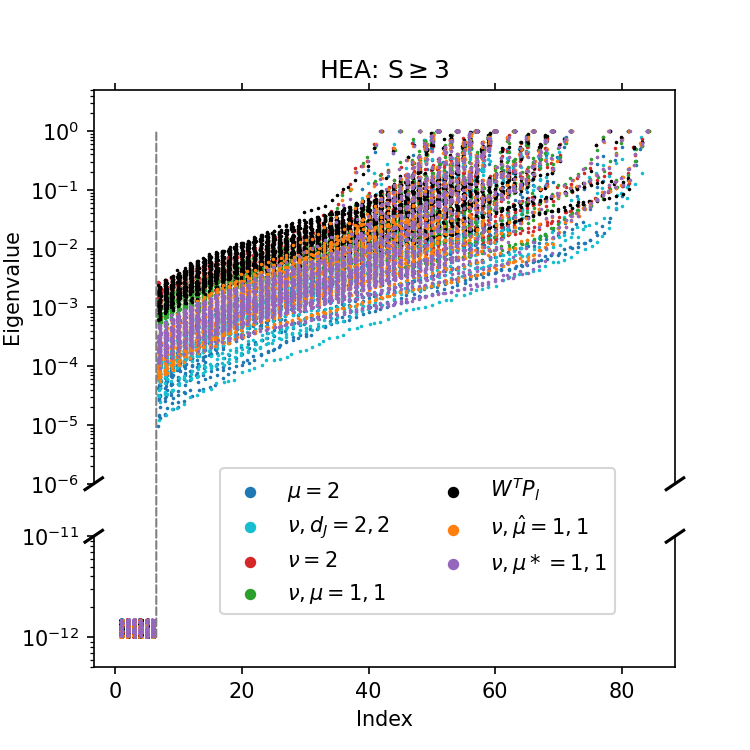}}
        {\includegraphics[width=0.95\textwidth]{example-image-a}}
    \end{minipage}
    \hfill
   \begin{minipage}{0.48\textwidth}
        \ifthenelse{\boolean{show_figs}}
        {\includegraphics[width=0.99\textwidth]{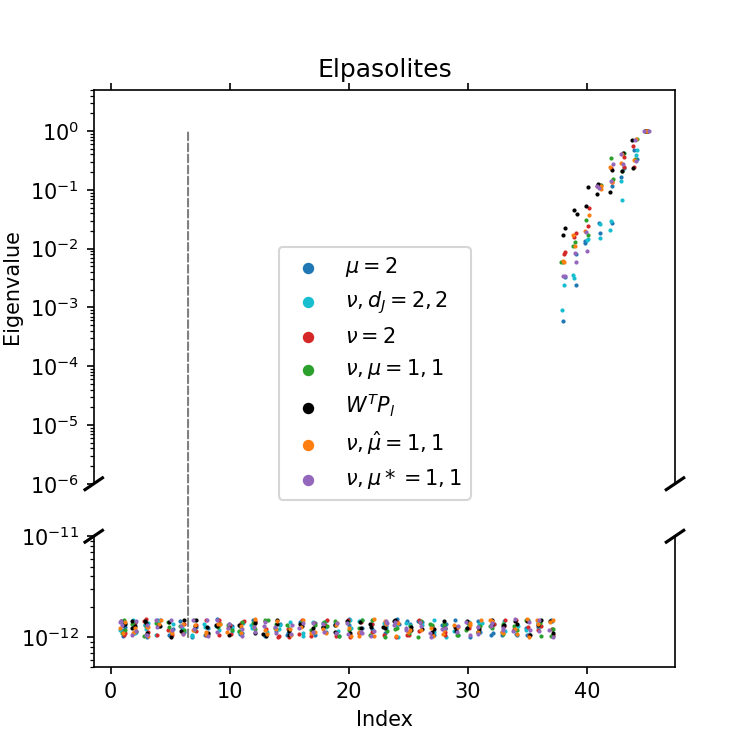}}
        {\includegraphics[width=0.95\textwidth]{example-image-a}}
    \end{minipage}
    \hfill
    \caption{The eigenvalues of the sensitivity matrix are shown for 50 randomly chosen liquid environments from the HEA dataset (left) and 3 randomly chosen environments from the elpasolite dataset (right). Values below $10^{-12}$ have been increased to this value and have had a small level of noise added to the y coordinate. All points have had random noise in the range $[-0.2, 0.2]$ added to the x coordinate to better see the data. The grey dashed line marks the 6 smallest eigenvalues, which are all zero to within numerical tolerance}
    \label{fig:sensitivity_overlay}
\end{figure*}

\begin{figure*}[htp]
  \ifthenelse{\boolean{show_figs}}
  {\includegraphics[width=0.95\textwidth]{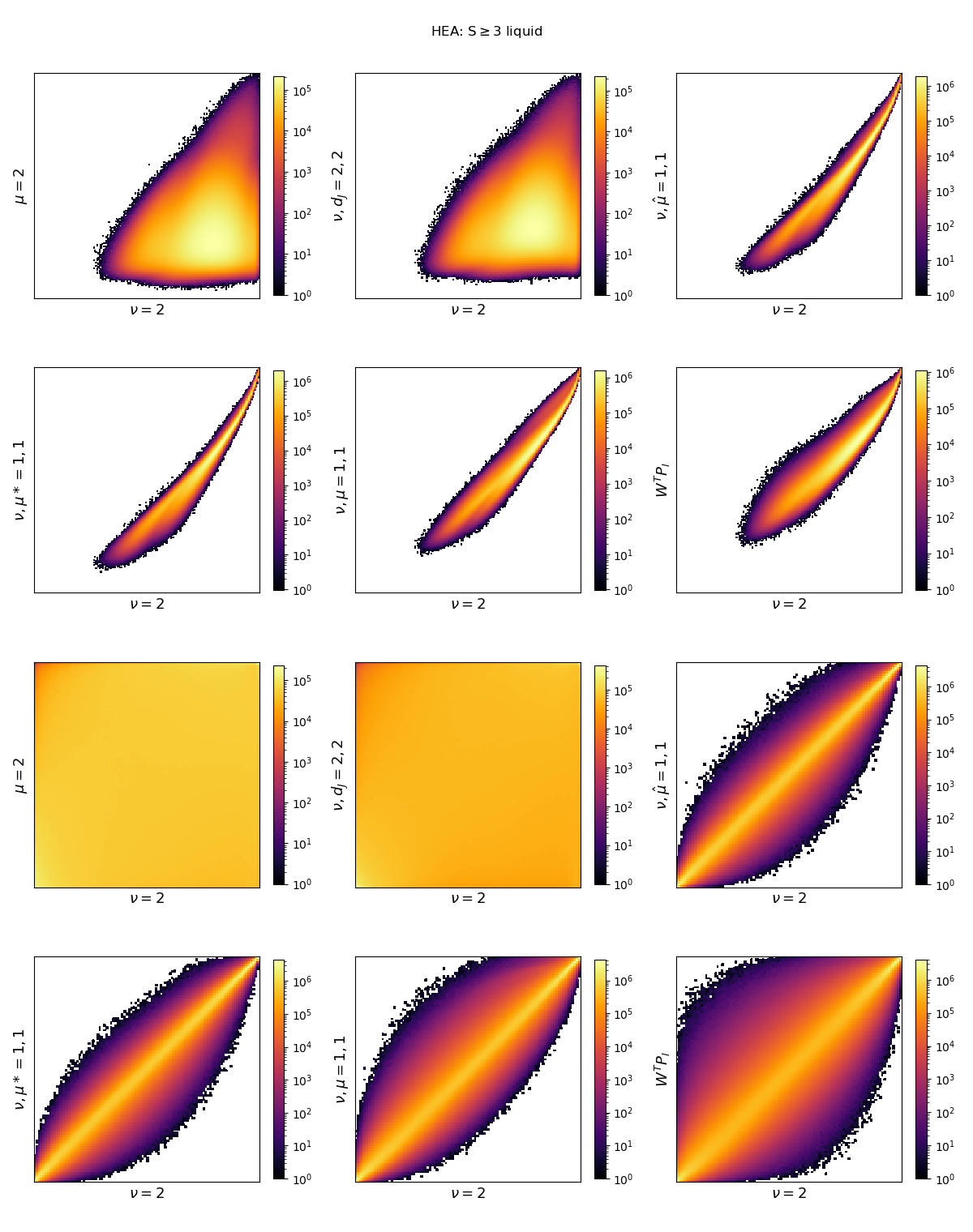}}
  {\includegraphics[width=0.95\textwidth]{example-image-a}}
  \caption{The top two rows show the distance-distance correlations between all pairs of liquid environments in the HEA dataset. The distance according to $\nu=2$ is shown on the $x$-axis whilst the distance according to the compressed descriptor is shown on the $y$-axis. For the bottom two rows the distances have been ranked, per environment, and the rank-rank correlation is shown. All descriptors were computed using $r_{cut}=5$ \AA, $n_{max}=12, l_{max}=9$, $\sigma=0.4$ \AA\space and a cut-off transition width of 0.5 \AA. The contribution of the central atom to the density was set to zero to allow for meaningful comparisons between environments with differing central atoms. All distances have been scaled so that the largest distance according to each descriptor is 1. The zero distances between environments and themselves have been excluded.}
  \label{Sfig:HEA_correlation}
\end{figure*}

\end{document}